# SPINEX-TimeSeries: Similarity-based Predictions with Explainable Neighbors Exploration for Time Series and Forecasting Problems

Ahmed Z. Naser[1], M.Z. Naser[2,3]
[1]Department of Mechanical Engineering, University of Manitoba, Canada, E-mail: a.naser@umanitoba.ca
[2]School of Civil & Environmental Engineering and Earth Sciences (SCEEES), Clemson University, USA
[3]Artificial Intelligence Research Institute for Science and Engineering (AIRISE), Clemson University, USA
E-mail: mznaser@clemson.edu, Website: www.mznaser.com

## Abstract

This paper introduces a new addition to the SPINEX (Similarity-based Predictions with Explainable Neighbors Exploration) family, tailored specifically for time series and forecasting analysis. This new algorithm leverages the concept of similarity and higher-order temporal interactions across multiple time scales to enhance predictive accuracy and interpretability in forecasting. To evaluate the effectiveness of SPINEX, we present comprehensive benchmarking experiments comparing it against 18 algorithms and across 49 synthetic and real datasets characterized by varying trends, seasonality, and noise levels. Our performance assessment focused on forecasting accuracy and computational efficiency. Our findings reveal that SPINEX consistently ranks among the top 5 performers in forecasting precision and has a superior ability to handle complex temporal dynamics compared to commonly adopted algorithms. Moreover, the algorithm's explainability features, Pareto efficiency, and medium complexity (on the order of O(log n)) are demonstrated through detailed visualizations to enhance the prediction and decision-making process. We note that integrating similarity-based concepts opens new avenues for research in predictive analytics, promising more accurate and transparent decision making.

## 1.0 Introduction

Time series analysis involves the study of data collected or recorded sequentially over time to extract meaningful patterns, trends, and insights [1]. Such a temporal ordering of data distinguishes time series from cross-sectional data and necessitates specialized techniques to understand the underlying mechanisms that generate the observed data and to forecast future values based on historical patterns. In other words, time series analysis inherently focuses on temporal data, with applications aiming to understand past behaviors, predict future trends, and identify cyclical patterns as well as anomalies. Such applications can span diverse domains, from financial forecasting to environmental modeling, etc. [2].

Given that a core component of time series analysis builds on forecasting future predictions from past trends/responses, the concept of similarity in time series then arises [3]. Similarity in this context refers to the degree of resemblance or correspondence between different segments of a single time series or between multiple time series [4]. The quantification of similarity enables researchers and practitioners to identify recurring patterns, classify time series into groups with similar characteristics, and detect deviations from expected behavior. In a way, similarity may enable the establishment of a notion of "normal" behavior, which can then help identify instances that exhibit low similarity to these reference patterns (i.e., anomalies). Thus, the notion of similarity can be particularly significant in pattern recognition, anomaly detection, and clustering [5].

In parallel, this concept is not only about finding sequences that look alike but also aims to understand temporal alignment and variation between data points. For example, two economic time series might exhibit similar patterns of growth and recession cycles but shift in time or at different scales. Accurately measuring such similarities involves techniques that consider alignment and scaling of time data instead of standard distance metrics [6]. This can be apparent in the case of traditional metrics like Euclidean and Manhattan distances, as these quantify similarity by measuring the distances between points in a time series. However, these measures often fall short of capturing the dynamic characteristics of time series data, such as trends and seasonality [7]. The same measures also tend to be sensitive to small fluctuations, which can be misleading in a temporal context where trends and cycles play a significant role [8].

As one can see, unlike static data, where similarity can often be measured using straightforward distance metrics, time series data presents additional complexities stemming from the observations' temporal nature. Consequently, specialized similarity measures have been developed to address these challenges, each with its own strengths and limitations. For instance, Dynamic Time Warping (DTW) allows elastic transformations of the time [9]. DTW also allows for non-linear alignment of time series and hence can accommodate differences in speed and duration of patterns. Thus, DTW can align two sequences in a way that minimizes their overall distance. This method is particularly effective in dealing with time series that are similar in shape but vary in speed or timing of events where temporal alignment is fundamental, such as speech recognition and bioinformatics [10,11].

Another similarity-based concept is the Longest Common Subsequence (LCSS), which measures the similarity between two sequences by identifying the longest subsequence present in both sequences without altering the order of elements [12]. LCSS can be robust to noise and occlusions and is particularly useful in real-time series applications where missing values may occur. Another approach to quantifying similarity in time series is through the use of feature-based methods and correlation measures. The former methods involve extracting relevant features or summary statistics from the time series and comparing these derived representations rather than the raw data points, including statistical moments, frequency domain characteristics, and model parameters. The latter measures, such as Pearson's or Spearman's correlation coefficients etc., capture the degree of relationship/association between time series. Both techniques have proven useful yet may fail to capture non-linear relationships and are sensitive to outliers and phase differences [13,14].

More recently, machine learning (ML) advancements have introduced new models for assessing similarity in time series data. Techniques such as Siamese and triplet networks learn similarity metrics directly from data, potentially capturing complex, non-linear relationships that traditional approaches might miss [15]. Furthermore, methods like K-Nearest Neighbors (KNN) for time series rely on identifying similar historical patterns to make predictions about future values. More sophisticated approaches, such as Long Short-Term Memory (LSTM) networks, can now implicitly learn to recognize and utilize similar patterns in their internal representations [16]. A key component that remains opaque in ML-based methods is their blackbox nature and limited interpretability [17].

It can then be inferred that the concept of similarity, as well as explainability, can be thought of as elemental to forecasting tasks. Thus, this paper presents the development of the time series variant to the SPINEX (Similarity-based Predictions with Explainable Neighbors Exploration) family. This variant builds upon the concept of similarity and explainability between similarly-identify neighbors and segments. As such, SPINEX hopes to bridge some of the existing challenges. In this study, we examine the algorithm's ability to perform on 49 diverse datasets compared to 18 commonly used algorithms.

## 2.0 Description of the SPINEX for time series and forecasting

### *2.1 General description*

SPINEX represents a unique approach to time series analysis and prediction. This algorithm integrates multiple techniques to deliver robust, adaptive, and interpretable time series forecasting. For example, at its core, SPINEX employs a multi-method similarity analysis, utilizing various measures such as cosine similarity, Euclidean distance, DTW, Pearson, and Spearman correlation. This ensemble approach enables a comprehensive assessment of segment similarities, capturing diverse aspects of time series behavior. A key feature of SPINEX is its adaptive window sizing mechanism, which adjusts based on data length, variability, and potential seasonality. This adaptability allows the algorithm to handle time series of varying lengths and characteristics effectively. Additionally, SPINEX implements time series cross-validation to provide robust performance estimates and assess model stability across different time periods.

SPINEX's multi-level analysis capability provides a hierarchical view of time series patterns to enable robustness to different scales of temporal dependencies. Furthermore, SPINEX incorporates a dynamic thresholding technique for anomaly detection and forecasting validation. This method adjusts the similarity threshold based on the recent performance of the predictions and the distribution of similarity scores, which further enhances the algorithm's flexibility and responsiveness to changing patterns in the data. The same can be crucial for understanding outliers and potential regime changes in the data. Thus, SPINEX effectively identifies the most relevant segments for making predictions to improve the reliability of the forecast. In cases where accessible similarity-based prediction is not feasible, the algorithm switches to a fallback prediction method, which includes trend extraction, multiple seasonality detection, non-linear trend modeling, and anomaly-aware residual prediction with confidence intervals.

As mentioned above, SPINEX focuses on explainability by offering detailed results on the most similar historical segments, their contributions to the prediction, and visualizations of the nearest neighboring similar segments. This enhances understanding of the prediction process and the underlying patterns in the data. Computational efficiency within SPINEX is achieved through the use of techniques like numba for just-in-time compilation and caching mechanisms.

*2.2 Detailed description*

A more detailed description of SPINEX's methods and functions is provided herein.

### Initialization (__init__)

The __init__ method initializes and sets up the operational parameters of SPINEX. The method signature is as follows:

```
def __init__(self, data, window_size=None, forecast_horizon=1, similarity_methods=None,
                 dynamic_window=True, multi_level=True, dynamic_threshold=True):
```

More specifically:

- **data:** The input time series data, converted into a NumPy array for efficient numerical operations.
- **window_size:** Determines the length of the segments to be compared. If not specified, it is set to the greater of 10 or one-tenth of the data length. This parameter can be dynamically adjusted based on the data's volatility.
- **forecast_horizon:** Specifies how far into the future predictions are made. By default, it is the smaller of the provided value and one-tenth of the data length.
- **similarity_methods**: A list of methods used to compute similarity between segments. Defaults to ['cosine', 'euclidean', 'dtw'] if not specified.
- **dynamic_window:** Enables or disables dynamic adjustment of the window size based on data characteristics.
- **multi_level:** Allows the use of multiple window sizes in the analysis to capture different scales of patterns.
- **dynamic_threshold:** Enables adaptive thresholding in the similarity calculations to improve forecast reliability.
- Additionally, the class uses caching mechanisms (similarity_cache and segments_cache) to store computed results for re-use to optimize performance for large datasets.

### Method: Similarity Measures

This method offers several methods to compute the similarity between time series segments:

- **Cosine Similarity:**
    - Computes the cosine of the angle between two vectors and is defined as the dot product of the vectors divided by the product of their norms. This measure is effective in identifying the similarity in direction regardless of magnitude.
    - Equation: $similarity = \frac{X \cdot X^T}{\|X\| \|X^T\|}$
- **Correlation Similarity:**
    - Calculates the Pearson correlation coefficient matrix of the rows of X, providing a measure of linear relationships between segments.
    - Equation: $similarity = \mathrm{corrcoef(X)}$
- **Euclidean Similarity:**
    - Uses the Euclidean distance to compute similarity by applying a transformation that inversely relates distance to similarity.
    - Equation: $similarity = \frac{1}{1+\sqrt{cdist(X, X, \mathrm{Euclidean})^2}}$
- **Spearman Similarity:**
    - Calculates the Spearman rank correlation between the columns of X, useful for capturing monotonic relationships between segments that may not necessarily be linear.
    - Equation: $similarity = \mathrm{spearmanr}(X^T)[0]$
- **Dynamic Time Warping (DTW) Similarity:**
    - Measures similarity based on the minimal distance that aligns two time series, accounting for shifts and distortions in time. In essence, DTW measures the similarity between two temporal

sequences, which may not be of the same length, by aligning their points to minimize the overall distance between them.

    - Equation: $similarity = \frac{1}{1+DTW(X[i],X[j])}$
        - $DTW\ (x, y) = \min(\text{cost} + \min(\text{dtw}_{matrix}[i-1, j], \text{dtw}_{matrix}[i, j-1], \text{dtw}_{matrix}[i-1, j-1])$

- Direction Similarity:
    - Calculates the direction similarity via the direction method to be discussed later on.

## Method: adjust_dynamic_parameters

This method adjusts the window size and similarity threshold based on the recent behavior of the time series and the algorithm's performance.

- Volatility-based Window Size Adjustment:
    - **Volatility Calculation:** First, this method calculates the volatility of the most recent portion of the data, defined as the standard deviation over the last data segments. The size of this segment is a maximum of 10 or one-tenth of the data length but not exceeding half the length of the data.
    - **Window Size Recalculation:** The window size is inversely adjusted based on the calculated volatility to respond to the data's fluctuating nature. If the volatility is low, a larger window size is used to smooth out noise and capture more extended patterns. If the volatility is high, the window size decreases, making the model more responsive to recent changes. A scaling factor controls this adjustment, clipped between 0.1 and 1.0 to prevent extreme values.
    - Equation: $window_size = max\ (MIN_WINDOW_SIZE, min\ (\frac{MAX_WINDOW_SIZE}{scale_factor}, MAX_WINDOW_SIZE))$
        - where scale_factor=clip(volatility,0.1,1.0).
- Threshold Adjustment:
    - **Error-based Adjustment:** If recent prediction errors are available, the threshold is adjusted based on these errors' mean and standard deviation to accommodate the model's predictive accuracy.
    - **Similarity Score-based Adjustment:** If recent similarity scores are tracked, the threshold is further adjusted to reflect the mean and variability in these scores. This dynamic threshold helps maintain the similarity measure's relevance under varying data conditions.
    - Equation: $threshold = mean_sim + std_sim + threshold_adjustment$

## Method: get_dynamic_threshold

This method computes an adaptive threshold for the similarity scores to decide which time series segments are considered similar enough to be relevant for predictions.

- **Basic Threshold Calculation:** Calculates a baseline threshold as the sum of the mean and standard deviation of the similarity scores. This method aims to keep only the most similar segments, thus ensuring that the predictions are based on the most relevant and recent data patterns.
- **Threshold Adjustment:** If fewer than five segments exceed this baseline threshold, indicating a potential over-tightening, the threshold is reduced to the 90th percentile of the scores to include more segments.
    - Equation: $adjusted_threshold = \begin{cases} percentile(similarities,\ 90)\ if\ abs\{s > base_threshold\} < 5 \\ base_threshold, otherwise \end{cases}$

## Method: adjusted_dtw_similarity

This method modifies the DTW similarity measure to be more forgiving by squaring the DTW distance before inversely transforming it into a similarity score. This adjustment makes the similarity measure less sensitive to small variations, emphasizing more significant patterns in the similarity assessment.

- Equation: $adjusted_scores = \frac{1}{1+\sqrt{dtw_scores}}$

### Method: plot_prediction

This method is designed to visualize the forecasting performance of the SPINEX model by plotting actual data alongside predicted values. This method serves as a tool for assessing the accuracy and relevance of the model's predictions.

### Method: extract_segments

This method prepares segments of the time series data for further analysis, such as computing similarities or making predictions such that:

- Dynamic Window Size: If no specific window size is provided, the method calculates an adaptive window size using the adaptive_window_size() method.
- Adjustment for Small Data: If the total data length is less than the determined window size, the window size is adjusted to half the data length to ensure at least some segmentation can be performed.
- Segmentation: Using np.lib.stride_tricks.sliding_window_view, the method creates overlapping segments of the specified window size from the time series data. This function efficiently generates a new view into the data array without copying the data.
- Normalization: Each segment is normalized by subtracting its mean and dividing by its standard deviation. This step standardizes the segments, mitigating the effect of different scales or baselines in the data and improving the comparability between segments.
    - Equation: $normalized_segments = \frac{segments - segment_means}{segment_stds + 1e-8}$
    - Here, 1e−8 is added to the standard deviations to prevent division by zero in the case of very uniform segments.

### Method: find_similar_segments

This method facilitates the identification of similar segments within the time series data, which is crucial for making accurate predictions.

- Multi-Level Analysis: Depending on the multi_level attribute, the method considers multiple window sizes for segmentation. These sizes include a smaller window (half the primary size), the primary window size itself, and a larger window (double the primary size or one-fourth the length of the data, whichever is smaller). This multi-scale approach allows the model to capture similarities at different granularities.
- Segment Extraction and Hashing: For each window size, the method extracts segments and computes a hash to uniquely identify them. This hash is used to cache the segments and avoid redundant calculations.
- Similarity Calculation: For each window size, the method computes similarity matrices using the specified methods (cosine, euclidean, dtw, etc.). If a large number of segments are detected (more than 500), DTW is skipped to avoid performance bottlenecks.
- Aggregation of Similarities: The method averages the similarities across different methods to get a composite similarity measure for each window size. These are then averaged across all window sizes to get the final measure of similarity between segments.
- Fallback Method: If no valid similarities are found (e.g., due to insufficient segments or errors in calculation), a fallback method based on autocorrelation is used.

### Method: fallback_similarity_method

This method provides a basic mechanism to calculate similarity based on autocorrelation when other methods fail or are not applicable due to data constraints.

### Method: analyze_segment_similarity

This method quantitatively assesses how similar a particular segment (indexed) is to the most recent segment in the time series.

- **Segment Extraction:** Both the target segment and a reference segment (usually the most recent one) are extracted.
- **Similarity Calculation:** The method calculates similarity scores using all available similarity methods, providing a detailed breakdown of how each method perceives the similarity.
- **Feature Contributions:** It calculates the absolute differences between the corresponding features of the two segments to determine which features contribute most to any dissimilarity.

### Method: get_nearest_neighbors

This method identifies the nearest neighbors of the most recent segment based on the computed similarities and can identify tasks for anomaly detection. After calculating similarities for all segments, it sorts these and picks the top k segments most similar to the latest segment, providing their indices and similarity scores.

### Method: detect_seasonality

This method is designed to identify seasonality within the time series data, which is crucial for understanding periodic patterns that could influence forecasting and other analytical tasks.

- **Autocorrelation Calculation:** The method calculates the autocorrelation (ACF) for the data up to a specified lag (max_lag). If max_lag is not specified, it defaults to half the length of the data.
- **Peak Detection:** The method then identifies ACF peaks, representing potential seasonal periods. Peaks are detected where the autocorrelation at a given lag is greater than its neighbors, indicating a repeating pattern.
- **Seasonality Inference:** If any peaks are detected, the first peak is assumed to represent the primary seasonal period, and its lag is returned. An empty list is returned if no peaks are detected, indicating no detectable seasonality.

### Method: detect_anomalies

This method identifies anomalies in the time series data by comparing the similarity of data segments to a dynamically determined threshold.

- **Segment Extraction and Similarity Calculation:** Segments of the data are extracted, and their similarities are computed.
- **Threshold Determination:** A threshold is set at a specified percentile (default is the 2nd percentile) of the similarity scores, identifying the least similar segments as potential anomalies.
- **Anomaly Identification:** Segments whose similarity scores fall below the threshold are marked as anomalies.
- Equation: $Threshold = percentile(similarities,\ threshold_percentile)$

### Method: fallback_prediction

This method provides a comprehensive mechanism for generating predictions when standard approaches are not feasible, utilizing multiple time series decomposition and modeling techniques.

- **Pre-checks:** It first ensures that there is sufficient data for prediction based on the specified number of points required.
- **Adaptive Window Sizing:** This step dynamically adjusts the window size for trend extraction based on minimizing the mean squared error (MSE) of the trend-subtracted data.
- **Trend Extraction:** Utilizes a moving average to smooth the data and extract the underlying trend.
- **Seasonality Detection:** Employs autocorrelation to identify potential seasonality periods and extract these seasonal components.

- **Residual Calculation:** The residuals (or unexplained components) are analyzed after removing the trend and seasonal components.
- **Anomaly Detection and Handling:** Anomalies in the residuals are identified and replaced with median values to stabilize the model.
- **Non-linear Trend Modeling:** Fits a polynomial model to predict the future trend based on past data.
- **Seasonal Component Prediction:** Projects the identified seasonal patterns into the future.
- **Residual Prediction:** Uses a weighted average approach to predict future residuals, incorporating confidence intervals to account for uncertainty.
- **Combination of Components:** The final prediction combines the trend, seasonal, and residual predictions to form a complete forecast.
- **Trend:** Extracted using a moving average filtered by convolution:
  - Equation: $trend\ = convolve(data, window)/window_size$
- **Seasonality:** Identified through peak detection in the autocorrelation function.
- **Residuals:** Calculated as data − trend
- **Confidence Intervals for Residuals:** Generated by assuming the residuals follow a normal distribution modulated by an exponential decay in influence.

## Method: tune_hyperparameters

This method optimizes the hyperparameters of the model, specifically focusing on the detection of seasonalities.

- **Iterative Testing**: The method iterates over a range of possible numbers of seasonalities to detect (from 1 to 4).
- **Prediction Generation**: For each candidate setting, it generates predictions using the fallback_prediction method.
- **Evaluation**: Calculates the MSE for each set of predictions compared to the actual data.
- **Selection of Optimal Parameter**: Identifies the number of seasonalities that result in the lowest MSE, suggesting the best fit for the data.

## Method: predict

This method combines various techniques to generate accurate forecasts based on the similarity of time series segments.

- **Dynamic Parameter Adjustment**: Initially, dynamic parameters such as window size and thresholds are adjusted based on recent data characteristics.
- **Similarity Assessment**: It calculates similarities between segments of the time series to identify patterns that can be used for forecasting.
- **Fallback Prediction**: If no significant similarities are found, it resorts to a fallback prediction method that uses more basic statistical methods.
- **Threshold Determination**: Determines a dynamic threshold for considering a segment significantly similar to the latest data, adjusting the threshold based on the distribution of similarity scores.
- **Valid Predictions Identification**: Identifies segments that meet the similarity threshold and ensures that they are within a valid range for making predictions.
- **Prediction Compilation**: Compiles predictions from multiple segments, weighted by their similarity scores, and adjusts them to align with the most recent actual data point.
- **Error Handling**: If any step fails, it defaults to the fallback prediction method.

## Method: update_recent_performance

This method updates the performance metrics of the model by recording the recent error and similarity scores, which are essential for monitoring and improving the model's accuracy over time.

- **Dynamic Parameter Adjustment**: Initially, dynamic parameters such as window size and thresholds are adjusted based on recent data characteristics.
- **Similarity Assessment**: It calculates similarities between segments of the time series to identify patterns that can be used for forecasting.

### Method: validate_prediction

This method evaluates the robustness of the model's predictions by using cross-validation, specifically time-series cross-validation, where the order of data points is preserved.

- **Setup:** Determines the number of splits for cross-validation based on available data, ensuring there are enough points for each training and testing set.
- **Cross-Validation:**
  - Single Split Handling: If there is insufficient data for multiple splits, perform a single train-test split.
  - Multiple Splits: Uses TimeSeriesSplit from scikit-learn to create training and testing segments. It ensures that predictions are based only on past data, respecting the temporal order.
  - Prediction and Evaluation: For each split, the model predicts future values based on the training set, and the predictions are evaluated using the evaluate_prediction method.
  - Aggregation of Results: The results from each split are aggregated to calculate average performance metrics across all splits.

### Method: get_explainability_results

This method provides insights into why certain predictions were made based on the similarity of time series segments.

- **Similarity Assessment:** The method first identifies similar segments by calculating and evaluating segment similarities.
- **Threshold Determination:** It dynamically determines a similarity threshold above which segments are considered significantly similar.
- **Top Segments Identification:** Segments surpassing the threshold are marked as key influencers. If no segments exceed the threshold, the top k segments based on similarity scores are selected.
- **Contribution Calculation:** Each top segment calculates how much each segment contributes to the predictions, using weighted averages based on their similarity scores.

### Method: analyze_and_plot_neighbors

This provides a deeper analysis of how and why certain segments are considered similar to the current segment, offering both visual and numerical insights.

- **Current and Neighbor Segment Extraction:** Similar to plot_nearest_neighbors, but with added analysis of segment similarities.
- **Similarity Analysis:** For each neighbor, it computes detailed similarity scores using various metrics.
- **Visualization and Reporting:** Each neighbor's segment and its similarity scores are plotted and displayed. This includes a breakdown of the scores for different similarity metrics and the identification of key features contributing to the similarities.
- **Similarity Scores:** Each neighbor's similarity to the current segment is quantified using methods like cosine, euclidean, and DTW similarities.
- **Feature Contributions:** Differences between segments are analyzed to pinpoint which specific elements (data points) contribute most to the observed similarities or discrepancies.

*<u>Additional functions for optimized clustering:</u>*

### Method: direction_accuracy

This method calculates the direction accuracy to compare the directional trends between two time series segments. Given two segments, segment1 and segment2, the following steps compute the direction accuracy:

- **Calculate the Differences:** compute the first-order differences of both segments such that:
    - Equation: $Diff1_i = segment1_{i+1} - segment1_i$ (with a similar approach for segment 2)
- **Determine the Direction**: Using the sign function sign(·), the direction of these differences can be calculated.
- **Equation**: $direction1_i = sign(Diff1_i)$ (with a similar approach for segment 2)
- **Compare Directions**: Compare the directional trends of the two segments by checking if the directions are equal at each time step
    - Equation: $match = \begin{cases} 1 \; if \; direction1_i = direction2_i \\ 0 \; otherwise \end{cases}$.

### Method: Entropy

The numba_sample_entropy method calculates the sample entropy of a sequence x, which is a measure of the complexity or the amount of regularity and unpredictability in time series data. This entropy is useful for determining the complexity of physiological time series signals.

- **Mathematical Representation**:
    - $Sample\ entropy = -\log \frac{A+1e-10}{B+1e-10}$

### Method: Hash Array (hash_array)

This static method generates a unique hash for a numpy array using MD5 to create keys for caching purposes, allowing efficient retrieval of previously computed results.

### Method: plot_anomalies

This method visualizes the anomalies detected.

### Method: plot_nearest_neighbors

This method visualizes the time series segments that are most similar to the most recent segment, facilitating an understanding of the model's decision-making process.

The complete class of SPINEX is shown below:

```
@jit(nopython=True)
def numba_dtw(x, y):
    n, m = len(x), len(y)
    dtw_matrix = np.zeros((n+1, m+1))
    for i in range(1, n+1):
        for j in range(1, m+1):
            cost = abs(x[i-1] - y[j-1])
            dtw_matrix[i, j] = cost + min(dtw_matrix[i-1, j], dtw_matrix[i, j-1], dtw_matrix[i-1, j-1])
    return dtw_matrix[n, m]

@jit(nopython=True)
def numba_dtw_similarity(X):
    n = X.shape[0]
    sim_matrix = np.zeros((n, n))
    for i in range(n):
        for j in range(i, n):
```

```
            dist = numba_dtw(X[i], X[j])
            sim_matrix[i, j] = sim_matrix[j, i] = 1 / (1 + dist)
    return sim_matrix

@jit(nopython=True)
def numba_sample_entropy(x, m=2, r=0.2):
    n = len(x)
    B = 0.0
    A = 0.0
    for i in range(n - m):
        for j in range(i + 1, n - m):
            matches = 0
            for k in range(m):
                if abs(x[i+k] - x[j+k]) <= r:
                    matches += 1
                else:
                    break
            if matches == m:
                B += 1
                if abs(x[i+m] - x[j+m]) <= r:
                    A += 1
    return -np.log((A + 1e-10) / (B + 1e-10))

def direction_accuracy(segment1, segment2):
    direction1 = np.sign(np.diff(segment1))
    direction2 = np.sign(np.diff(segment2))
    return np.mean(direction1 == direction2)

class SPINEX_Timeseries:
    def __init__(self, data, window_size=None, forecast_horizon=1, similarity_methods=None,
                dynamic_window=True, multi_level=True, dynamic_threshold=True):
        self.data = np.array(data)
        if window_size is None:
            self.window_size = max(10, len(data) // 10)
        else:
            self.window_size = min(window_size, len(data) // 2)
        self.forecast_horizon = min(forecast_horizon, len(data) // 10)
        self.forecast_horizon = forecast_horizon
        self.similarity_methods = similarity_methods if similarity_methods else ['cosine', 'euclidean', 'dtw']
        self.similarity_cache = {}
        self.dynamic_window = dynamic_window
        self.multi_level = multi_level
        self.dynamic_threshold = dynamic_threshold
        self.segments_cache = {}
        self.recent_errors = []
        self.recent_similarity_scores = []
        if self.dynamic_window:
            self.window_size = self.adaptive_window_size()
```

```
@staticmethod
def hash_array(arr):
    return hashlib.md5(arr.data.tobytes()).hexdigest()

@lru_cache(maxsize=128)
def get_similarity_matrix(self, method, segments_hash):
    if (segments_hash, method) in self.similarity_cache:
        return self.similarity_cache[(segments_hash, method)]
    segments = self.segments_cache[segments_hash]
    if method == 'cosine':
        similarity_matrix = self.cosine_similarity(segments)
    elif method == 'correlation':
        similarity_matrix = self.correlation_similarity(segments)
    elif method == 'euclidean':
        similarity_matrix = self.euclidean_similarity(segments)
    elif method == 'spearman':
        similarity_matrix = self.spearman_similarity(segments)
    elif method == 'dtw':
        similarity_matrix = numba_dtw_similarity(segments)
    elif method == 'direction':
        similarity_matrix = self.direction_similarity(segments)
    else:
        raise ValueError(f"Invalid similarity method: {method}")
    self.similarity_cache[(segments_hash, method)] = similarity_matrix
    return similarity_matrix

@staticmethod
def cosine_similarity(X):
    norm = np.linalg.norm(X, axis=1)
    return np.dot(X, X.T) / np.outer(norm, norm)

@staticmethod
def correlation_similarity(X):
    return np.corrcoef(X)

@staticmethod
def euclidean_similarity(X):
    sq_dists = cdist(X, X, metric='euclidean')**2
    return 1 / (1 + np.sqrt(sq_dists))

@staticmethod
def spearman_similarity(X):
    return spearmanr(X.T)[0]

def adjust_dynamic_parameters(self):
    MIN_WINDOW_SIZE = 10
    MAX_WINDOW_SIZE = len(self.data) // 2
    BASELINE_WINDOW_SIZE = max(MIN_WINDOW_SIZE, len(self.data) // 10)
    if len(self.data) > BASELINE_WINDOW_SIZE:
```

```
        volatility = np.std(self.data[-BASELINE_WINDOW_SIZE:])
    else:
        volatility = np.std(self.data)
    scale_factor = np.clip(volatility, 0.1, 1.0)  # Limiting scale factor to avoid extreme values
    self.window_size = int(MAX_WINDOW_SIZE / scale_factor)
    self.window_size = max(MIN_WINDOW_SIZE, min(self.window_size, MAX_WINDOW_SIZE))
    if hasattr(self, 'recent_errors'):
        recent_error_mean = np.mean(self.recent_errors)
        recent_error_std = np.std(self.recent_errors)
        threshold_adjustment = recent_error_mean + recent_error_std
    else:
        threshold_adjustment = 0
    if hasattr(self, 'recent_similarity_scores') and self.recent_similarity_scores:
        mean_sim = np.mean(self.recent_similarity_scores)
        std_sim = np.std(self.recent_similarity_scores)
        self.threshold = mean_sim + std_sim + threshold_adjustment
    else:
        self.threshold = 0.5  # Default threshold if no recent similarities are recorded
    print(f"Adjusted Window Size: {self.window_size}, Threshold: {self.threshold}")

def get_dynamic_threshold(self, similarities):
    if self.dynamic_threshold:
        mean_sim = np.mean(similarities)
        std_sim = np.std(similarities)
        base_threshold = mean_sim + std_sim
        if len(similarities[similarities > base_threshold]) < 5:
            # If less than 5 indices are above threshold, reduce it to include more indices
            adjusted_threshold = np.percentile(similarities, 90)  # Adjusting percentile upward
        else:
            adjusted_threshold = base_threshold
        print(f"Dynamic Threshold Adjusted: {adjusted_threshold}")
        return adjusted_threshold
    else:
        return np.percentile(similarities, 95)

def adjusted_dtw_similarity(self, X):
    dtw_scores = numba_dtw_similarity(X)
    adjusted_scores = 1 / (1 + np.sqrt(dtw_scores))  # Squaring DTW scores for more lenience
    return adjusted_scores

def plot_prediction(self):
    predicted_values = self.predict()
    if predicted_values.size > 0:
        prediction_start_index = len(self.data) - self.forecast_horizon
        plt.figure(figsize=(12, 6))
        plt.plot(self.data, label='Actual Time Series', color='blue')
        plt.plot(np.arange(prediction_start_index, len(self.data)),
                 self.data[prediction_start_index:], label='Actual (Prediction Window)', color='green')
        plt.plot(np.arange(prediction_start_index, len(self.data)),
```

```
                predicted_values, label='Predicted', color='red', linestyle='--')
        plt.title('Time Series Prediction Comparison')
        plt.xlabel('Time Index')
        plt.ylabel('Values')
        plt.legend()
        plt.show()
    else:
        print("No valid predictions could be made.")

@lru_cache(maxsize=32)
def extract_segments(self, window_size=None):
    if window_size is None:
        window_size = self.adaptive_window_size()
    data_length = len(self.data)
    if data_length < window_size:
        print(f"Data length ({data_length}) is less than window size ({window_size}). Adjusting window size.")
        window_size = data_length // 2  # Use half of data length as window size
    n = data_length - window_size + 1
    if n <= 1:
        return np.array([self.data[-window_size:]])
    segments = np.lib.stride_tricks.sliding_window_view(self.data, window_size)
    segment_means = np.mean(segments, axis=1)
    segment_stds = np.std(segments, axis=1)
    normalized_segments = (segments - segment_means[:, np.newaxis]) / (segment_stds[:, np.newaxis] + 1e-8)
    return normalized_segments
def find_similar_segments(self):
    window_sizes = [self.window_size]
    if self.multi_level:
         window_sizes = [max(2, self.window_size // 2)] + window_sizes + [min(len(self.data) // 4, self.window_size *
2)]
    all_similarities = []
    for w_size in window_sizes:
        segments = self.extract_segments(w_size)
        if len(segments) < 2:
            print(f"Not enough segments for window size {w_size}, skipping.")
            continue
        segments_hash = self.hash_array(segments)
        self.segments_cache[segments_hash] = segments
        method_similarities = []
        for method in self.similarity_methods:
            if method == 'dtw' and len(segments) > 500:
                print(f"DTW skipped for large dataset with {len(segments)} segments.")
                continue
            try:
                sim_matrix = self.get_similarity_matrix(method, segments_hash)
                if sim_matrix.ndim > 1:
                    method_similarities.append(sim_matrix[-1, :-1])
                else:
                    method_similarities.append(sim_matrix[:-1])
```

```
            except Exception as e:
                print(f"Error calculating similarity for method {method}: {str(e)}")
        if not method_similarities:
            print(f"No valid similarity methods for window size {w_size}, skipping.")
            continue
        min_length = min(len(sim) for sim in method_similarities)
        method_similarities = [sim[-min_length:] for sim in method_similarities]
        method_similarities_array = np.array(method_similarities)
        overall_similarity = np.nanmean(method_similarities_array, axis=0)
        all_similarities.append(overall_similarity)
    if not all_similarities:
        print("No similarities found for any window size. Using fallback similarity.")
        return self.fallback_similarity_method()
    min_length = min(len(s) for s in all_similarities)
    all_similarities = [s[-min_length:] for s in all_similarities]
    all_similarities_array = np.array(all_similarities)
    combined_similarities = np.nanmean(all_similarities_array, axis=0)
    return combined_similarities

def fallback_similarity_method(self):
    # Simple autocorrelation-based similarity
    acf = np.correlate(self.data, self.data, mode='full')[len(self.data)-1:]
    return acf / acf[0]  # Normalize

def analyze_segment_similarity(self, segment_index):
    current_segment = self.extract_segments(self.window_size)[-1]
    historical_segment = self.extract_segments(self.window_size)[segment_index]
    similarity_scores = {}
    for method in self.similarity_methods:
        if method == 'cosine':
                score = np.dot(current_segment, historical_segment) / (np.linalg.norm(current_segment) *
np.linalg.norm(historical_segment))
        elif method == 'euclidean':
            score = 1 / (1 + np.linalg.norm(current_segment - historical_segment))
        elif method == 'dtw':
            score = 1 / (1 + numba_dtw(current_segment, historical_segment))  # Use the global function
        similarity_scores[method] = score
    feature_contributions = np.abs(current_segment - historical_segment)
    top_contributing_features = np.argsort(feature_contributions)[::-1][:5]
    return {
        'similarity_scores': similarity_scores,
        'top_contributing_features': top_contributing_features.tolist(),
        'feature_contributions': feature_contributions.tolist()
    }

def get_nearest_neighbors(self, k=5):
    similarities = self.find_similar_segments()
    nearest_indices = np.argsort(similarities)[::-1][:k]
    return [(idx, similarities[idx]) for idx in nearest_indices]
```

```
def dtw_similarity(self, X):
    return numba_dtw_similarity(X)  # Use the global function

def adaptive_window_size(self):
    data_length = len(self.data)
    if data_length < 100:
        base_window = max(2, data_length // 20)
    elif data_length < 1000:
        base_window = max(5, data_length // 40)
    else:
        base_window = max(25, data_length // 80)
    potential_seasons = self.detect_seasonality()
    variability = np.std(self.data) / (np.mean(self.data) + 1e-8)
    if potential_seasons:
        window = min(max(potential_seasons), base_window)
    else:
        window = int(base_window * (1 + variability))
    return max(2, min(window, data_length // 8))  # Ensure window is at most 1/8 of data length

def detect_seasonality(self, max_lag=None):
    if max_lag is None:
        max_lag = len(self.data) // 2
    acf = np.correlate(self.data, self.data, mode='full')[-max_lag:]
    peaks = np.where((acf[1:-1] > acf[:-2]) & (acf[1:-1] > acf[2:]))[0] + 1
    if len(peaks) > 0:
        return [int(peaks[0])]  # Return a list with the first peak
    return []  # Return an empty list if no peaks found
def detect_anomalies(self, threshold_percentile=2):
    segments = self.extract_segments(self.window_size)
    similarities = self.find_similar_segments()
    threshold = np.percentile(similarities, threshold_percentile)
    anomaly_indices = np.where(similarities < threshold)[0]
    anomalies = []
    for idx in anomaly_indices:
        start = idx
        end = idx + self.window_size
        anomalies.append({
            'start_index': start,
            'end_index': end,
            'segment': self.data[start:end].tolist(),
            'similarity_score': similarities[idx]
        })
    return anomalies, threshold

def plot_anomalies(self, threshold_percentile=5):
    anomalies, threshold = self.detect_anomalies(threshold_percentile)
    plt.figure(figsize=(12, 6))
    plt.plot(self.data, label='Time Series', color='blue')
```

```
    for anomaly in anomalies:
        plt.axvspan(anomaly['start_index'], anomaly['end_index'], color='red', alpha=0.3)
    plt.title(f'Time Series with Detected Anomalies (Threshold: {threshold:.4f})')
    plt.xlabel('Time Index')
    plt.ylabel('Values')
    plt.legend()
    if not anomalies:
        plt.text(0.5, 0.5, 'No anomalies detected', horizontalalignment='center',
                verticalalignment='center', transform=plt.gca().transAxes)
    else:
        print(f"Detected {len(anomalies)} anomalies")
    plt.show()

    similarities = self.find_similar_segments()
    print(f"Similarity score range: {similarities.min():.4f} to {similarities.max():.4f}")
    print(f"Similarity score mean: {similarities.mean():.4f}")
    print(f"Similarity score median: {np.median(similarities):.4f}")
    print(f"Anomaly threshold: {threshold:.4f}")

def calculate_mean_squared_error(self, actual, predicted):
    return np.mean((actual - predicted) ** 2)

def calculate_basic_similarity(self, actual, predicted):
    # Ensuring that neither actual nor predicted are empty to avoid runtime errors
    if actual.size == 0 or predicted.size == 0:
        return np.nan
    correlation = np.corrcoef(actual, predicted)[0, 1]
    return correlation

def fallback_prediction(self, num_points):
    if len(self.data) < num_points * 2:
        raise ValueError("Insufficient data for prediction")
    def adaptive_window(data):
        def mse(window):
            trend = extract_trend(data, int(window))
            return np.mean((data[int(window)-1:] - trend)**2)
        result = minimize_scalar(mse, bounds=(10, len(data)//2), method='bounded')
        return int(result.x)

    def extract_trend(data, window_size):
        return np.convolve(data, np.ones(window_size), 'valid') / window_size

    def detect_seasonalities(data, max_period, num_seasons=2):
        correlations = [np.corrcoef(data[:-i], data[i:])[0, 1] for i in range(1, max_period)]
        seasons = []
        for _ in range(num_seasons):
            if len(correlations) > 0:
                season = np.argmax(correlations) + 1
                seasons.append(season)
```

```
            correlations[season-1] = -1  # Remove detected season
    return seasons

def model_nonlinear_trend(data, x):
    coeffs = np.polyfit(x, data, 3)
    return np.poly1d(coeffs)

def detect_anomalies(data, threshold=3):
    mean = np.mean(data)
    std = np.std(data)
    return np.abs(data - mean) > threshold * std
window_size = adaptive_window(self.data)
trend = extract_trend(self.data, window_size)
detrended = self.data[window_size-1:] - trend
seasonality_periods = detect_seasonalities(detrended, num_points)
seasonals = []
for period in seasonality_periods:
    seasonal = np.zeros(period)
    for i in range(period):
        seasonal[i] = np.mean(detrended[i::period])
    seasonals.append(seasonal)
combined_seasonal = np.zeros_like(detrended)
for seasonal in seasonals:
    combined_seasonal += np.tile(seasonal, len(detrended) // len(seasonal) + 1)[:len(detrended)]
residuals = detrended - combined_seasonal[:len(detrended)]
anomalies = detect_anomalies(residuals)
cleaned_residuals = residuals.copy()
cleaned_residuals[anomalies] = np.median(residuals)
x = np.arange(len(self.data))
trend_model = model_nonlinear_trend(self.data, x)
future_x = np.arange(len(self.data), len(self.data) + self.forecast_horizon)
future_trend = trend_model(future_x)
future_seasonal = np.zeros(self.forecast_horizon)
for seasonal in seasonals:
    future_seasonal += np.tile(seasonal, self.forecast_horizon // len(seasonal) + 1)[:self.forecast_horizon]

def predict_residuals_with_ci(residuals, horizon, confidence=0.95):
    weights = np.exp(np.linspace(-1, 0, len(residuals)))
    weighted_mean = np.sum(residuals * weights) / np.sum(weights)
    weighted_std = np.sqrt(np.sum(weights * (residuals - weighted_mean)**2) / np.sum(weights))
    predictions = np.random.normal(weighted_mean, weighted_std, (1000, horizon))
    mean_prediction = np.mean(predictions, axis=0)
    ci_lower = np.percentile(predictions, (1 - confidence) / 2 * 100, axis=0)
    ci_upper = np.percentile(predictions, (1 + confidence) / 2 * 100, axis=0)
    return mean_prediction, ci_lower, ci_upper
future_residuals, ci_lower, ci_upper = predict_residuals_with_ci(cleaned_residuals, self.forecast_horizon)
predictions = future_trend + future_seasonal + future_residuals
ci_lower += future_trend + future_seasonal
ci_upper += future_trend + future_seasonal
```

```
    return predictions, ci_lower, ci_upper

def tune_hyperparameters(self):
    # Example: tune the number of seasonalities to detect
    best_num_seasons = 1
    best_mse = float('inf')
    for num_seasons in range(1, 5):
        predictions, _, _ = self.fallback_prediction(num_points=20)
        mse = np.mean((self.data[-len(predictions):] - predictions)**2)
        if mse < best_mse:
            best_mse = mse
            best_num_seasons = num_seasons
    return {'num_seasons': best_num_seasons}

def predict(self):
    self.adjust_dynamic_parameters()
    try:
        similarities = self.find_similar_segments()
        if len(similarities) == 0:
            print("No similarities found. Using fallback prediction.")
            return self.fallback_prediction(self.forecast_horizon)[0]
        threshold = self.get_dynamic_threshold(similarities)
        valid_indices = []
        for percentile in range(95, 70, -5):  # Start at 95th percentile, go down to 70th
            top_indices = np.where(similarities > np.percentile(similarities, percentile))[0]
            valid_indices = top_indices[top_indices + self.window_size + self.forecast_horizon <= len(self.data)]
            if len(valid_indices) >= 3:
                break
        if len(valid_indices) == 0:
            print("No valid indices found. Using fallback prediction.")
            return self.fallback_prediction(self.forecast_horizon)[0]
        predictions = []
        weights = []
        for idx in valid_indices:
            start = idx + self.window_size
            end = start + self.forecast_horizon
            if end <= len(self.data):
                segment = self.data[start:end]
                predictions.append(segment)
                weights.append(similarities[idx])
        if predictions:
            min_length = min(len(p) for p in predictions)
            predictions = [p[:min_length] for p in predictions]
            predictions = np.array(predictions)
            weights = np.array(weights)
            last_actual = self.data[-1]
            for i in range(len(predictions)):
                shift = last_actual - predictions[i][0]
                predictions[i] += shift
```

```
            predicted_values = np.average(predictions, axis=0, weights=weights)
        else:
            print("No valid predictions. Using fallback prediction.")
            predicted_values = self.fallback_prediction(self.forecast_horizon)[0]
    except Exception as e:
        print(f"Error in predict: {str(e)}")
        predicted_values = self.fallback_prediction(self.forecast_horizon)[0]  # Return only predictions, not CI
    if predicted_values.size > 0:
        actual_values = self.data[-len(predicted_values):]
        prediction_error = self.calculate_mean_squared_error(actual_values, predicted_values)
        recent_similarity_score = self.calculate_basic_similarity(actual_values, predicted_values)
        self.update_recent_performance(prediction_error, recent_similarity_score)
    else:
        self.update_recent_performance(np.nan, np.nan)
    return predicted_values

def update_recent_performance(self, new_error, new_similarity_score):
    self.recent_errors.append(new_error)
    self.recent_similarity_scores.append(new_similarity_score)
    # Optionally, trim these lists to avoid unlimited growth
    self.recent_errors = self.recent_errors[-100:]  # Keep the last 100 records
    self.recent_similarity_scores = self.recent_similarity_scores[-100:]

def evaluate_prediction(self, actual, predicted):
    if len(actual) != len(predicted):
        raise ValueError("Actual and predicted arrays must have the same length.")
    if len(actual) == 0:
        return {metric: np.nan for metric in ['MSE', 'MAE', 'RMSE', 'MAPE', 'SMAPE', 'R-squared', 'Direction Accuracy', 'Theil\'s U']}
    mse = np.mean((actual - predicted) ** 2)
    mae = np.mean(np.abs(actual - predicted))
    rmse = np.sqrt(mse)
    mape = np.mean(np.abs((actual - predicted) / (actual + 1e-8))) * 100
    smape = np.mean(2 * np.abs(predicted - actual) / (np.abs(actual) + np.abs(predicted) + 1e-8)) * 100
    r2 = r2_score(actual, predicted)
    direction_actual = np.sign(np.diff(actual))
    direction_pred = np.sign(np.diff(predicted))
    direction_accuracy = np.mean(direction_actual == direction_pred) * 100
    actual_changes = np.diff(actual)
    predicted_changes = np.diff(predicted)
    theil_u = np.sqrt(np.sum(predicted_changes**2) / np.sum(actual_changes**2)) if np.sum(actual_changes**2) != 0 else np.nan
    return {
        'MSE': mse, 'MAE': mae, 'RMSE': rmse, 'MAPE': mape, 'SMAPE': smape,
        'R-squared': r2, 'Direction Accuracy': direction_accuracy, 'Theil\'s U': theil_u
    }

def validate_prediction(self, splits=3):
        n_samples = len(self.data)
```

```
        max_splits = (n_samples - self.window_size) // self.forecast_horizon
        splits = min(splits, max_splits)
        if splits < 2:
            print("Warning: Not enough data for multiple splits. Performing single train-test split.")
            train_size = int(0.8 * n_samples)
            train, test = self.data[:train_size], self.data[train_size:]
            self.data = train
            self.similarity_cache = {}
            predicted = self.predict()
            if predicted.size > 0:
                actual = test[:len(predicted)]
                metrics = self.evaluate_prediction(actual, predicted)
                self.data = np.concatenate((train, test))  # Restore original data
                return metrics
            else:
                print("Insufficient data to make a prediction.")
                return None
        tscv = TimeSeriesSplit(n_splits=splits, test_size=self.forecast_horizon)
        errors = []
        for train_index, test_index in tscv.split(self.data):
            if len(train_index) < self.window_size:
                print(f"Warning: Train set too small for window size. Skipping split.")
                continue
            train, test = self.data[train_index], self.data[test_index]
            original_data = self.data
            self.data = train
            self.similarity_cache = {}
            predicted = self.predict()
            if predicted.size > 0:
                actual = test[:len(predicted)]
                metrics = self.evaluate_prediction(actual, predicted)
                errors.append(metrics)
            else:
                print("Insufficient data to predict for this split.")
            self.data = original_data
        if errors:
            avg_metrics = {metric: np.mean([e[metric] for e in errors if metric in e]) for metric in errors[0]}
            return avg_metrics
        else:
            print("No valid predictions could be made across splits.")
            return None

    def get_explainability_results(self, top_k=5):
        similarities = self.find_similar_segments()
        threshold = self.get_dynamic_threshold(similarities)
        top_indices = np.where(similarities > threshold)[0]
        if len(top_indices) == 0:
            top_indices = np.argsort(similarities)[-top_k:]
        results = {
```

```
            'top_similar_segments': top_indices.tolist(),
            'similarity_scores': similarities[top_indices].tolist(),
            'threshold': threshold,
            'segment_contributions': []
        }
        predictions = []
        valid_indices = []
        for idx in top_indices:
            start = idx + self.window_size
            if start + self.forecast_horizon <= len(self.data):
                predictions.append(self.data[start:start + self.forecast_horizon])
                valid_indices.append(idx)
        if not predictions:
            return results
        predictions = np.array(predictions)
        weights = similarities[valid_indices]
        weighted_predictions = predictions * weights[:, np.newaxis]
        for i, (index, score, prediction, contribution) in enumerate(zip(valid_indices, similarities[valid_indices],
predictions, weighted_predictions)):
            results['segment_contributions'].append({
                'segment_index': int(index),
                'similarity_score': float(score),
                'prediction': prediction.tolist(),
                'weighted_contribution': contribution.tolist(),
                'contribution_percentage': (contribution / np.sum(weighted_predictions, axis=0) * 100).tolist()
            })
        return results

    def plot_nearest_neighbors(self, k=5):
        current_segment = self.extract_segments(self.window_size)[-1]
        neighbors = self.get_nearest_neighbors(k)
        plt.figure(figsize=(15, 10))
        plt.subplot(k+1, 1, 1)
        plt.plot(current_segment, color='blue', label='Current Segment')
        plt.title('Current Segment')
        plt.legend()
        for i, (idx, similarity) in enumerate(neighbors, start=2):
            neighbor_segment = self.extract_segments(self.window_size)[idx]
            plt.subplot(k+1, 1, i)
            plt.plot(neighbor_segment, color='red', label=f'Neighbor {i-1}')
            plt.title(f'Neighbor {i-1} (Similarity: {similarity:.4f})')
            plt.legend()
        plt.tight_layout()
        plt.show()

    def analyze_and_plot_neighbors(self, k=5):
        current_segment = self.extract_segments(self.window_size)[-1]
        neighbors = self.get_nearest_neighbors(k)
        plt.figure(figsize=(20, 5*k))
```

```
    plt.subplot(k+1, 2, 1)
    plt.plot(current_segment, color='blue', label='Current Segment')
    plt.title('Current Segment')
    plt.legend()
    for i, (idx, overall_similarity) in enumerate(neighbors, start=1):
        neighbor_segment = self.extract_segments(self.window_size)[idx]
        analysis = self.analyze_segment_similarity(idx)
        plt.subplot(k+1, 2, 2*i+1)
        plt.plot(neighbor_segment, color='red', label=f'Neighbor {i}')
        plt.title(f'Neighbor {i} (Overall Similarity: {overall_similarity:.4f})')
        plt.legend()
        plt.subplot(k+1, 2, 2*i+2)
        methods = list(analysis['similarity_scores'].keys())
        scores = list(analysis['similarity_scores'].values())
        plt.bar(methods, scores)
        plt.title(f'Similarity Scores for Neighbor {i}')
        plt.ylim(0, 1)
        print(f"\nNeighbor {i} Analysis:")
        print(f"Overall Similarity: {overall_similarity:.4f}")
        print("Similarity Scores:")
        for method, score in analysis['similarity_scores'].items():
            print(f"  {method}: {score:.4f}")
        print("Top Contributing Features:", analysis['top_contributing_features'])
    plt.tight_layout()
    plt.show()
```

## 3.0 Description of benchmarking algorithms

We examined SPINEX against 18 commonly used time series forecasting algorithms, namely, ARIMA, SARIMA, ETS, Holt-Winters, Prophet, Theta, Simple Moving Average, VAR, Croston's Method, LSTM, Neural Networks, Gaussian Process Regression, KNN, SVR, Random Forest, XGBoost, Gradient Boosting, CatBoost, and Bagging. As one can see, the first nine algorithms are specifically designed for time series analysis, while the latter group consists of other ML algorithms that can be adapted for time series forecasting with appropriate feature engineering, as seen in [8,18–20]. Each of these algorithms is described in this section, where we showcase a brief historical background and algorithmic logic (with additional details being available in the cited original sources). Table 1 compares these algorithms with respect to their time series forecasting characteristics.

### *3.1 Algorithms specifically designed for time series analysis*

#### 3.1.1 Autoregressive Integrated Moving Average (ARIMA and SARIMA)

ARIMA (Autoregressive Integrated Moving Average) and its seasonal variant SARIMA were popularized by Box and Jenkins in the 1970s [21] – however, the concepts of Auto-Regressive and Moving Average models were introduced by Yule in 1926 and by Slutsky in 1937, respectively [22]. The ARIMA algorithm combines these concepts and components with differencing to handle non-stationary data. The ARIMA algorithm is particularly effective for univariate time series forecasting, while SARIMA extends this capability to series with seasonal patterns. The models are specified by three main parameters: p (order of the Auto-Regressive term), d (degree of differencing), and q (order of the Moving Average term), and SARIMA adds additional seasonal parameters. These algorithms are widely used due to their flexibility and ability to capture complex temporal dependencies. However, the algorithms assume linear relationships and, hence, may struggle with highly nonlinear patterns. Moreover, the selection of appropriate internal parameters can be challenging and often requires expert/domain knowledge or automated procedures [23].

#### 3.1.2 Croston's Method

Croston introduced this method in 1972 [24] as a specialized forecasting algorithm designed for intermittent demand patterns. This algorithm separates the time series into two components: the non-zero demand sizes and the intervals between non-zero demands. Each component is then forecasted separately using simple exponential smoothing, and the final forecast is obtained by dividing the demand size forecast by the interval forecast. This method is particularly useful in domains where demand occurs sporadically (such as that commonly seen in inventory management and spare parts forecasting) [25]. Croston's method assumes that the demand sizes and intervals are independent (which often introduces bias as this assumption may not always hold true). Several modifications of Croston's method have been proposed to address this main limitation [26,27].

#### 3.1.3 Error, Trend, Seasonality (ETS), and the Holt-Winters Method

ETS (Error, Trend, Seasonality) and Holt-Winters methods are exponential smoothing techniques that have evolved since their introduction by Brown and Holt in the 1950s [28,29]. These two methods decompose time series into components (level, trend, and seasonality) and use weighted averages of past observations to forecast future values. ETS provides a framework for selecting the most appropriate model based on the nature of the components (i.e., additive or multiplicative). Holt-Winters [30] is a specific implementation within the ETS family that has been modified to account for time series with both trend and seasonal components. This family of algorithms can be effective in handling a wide range of time series patterns. However, these algorithms may struggle with complex, non-linear relationships and can be sensitive to outliers [31].

#### 3.1.4 Long Short-Term Memory (LSTM)

The Long Short-Term Memory (LSTM) network was introduced by Hochreiter and Schmidhuber in 1997 [32] as a type of recurrent neural network. This network is designed to capture long-term dependencies in sequential data. LSTMs use a series of gates (input, forget, and output gates) to control the flow of information through the network, allowing them to selectively remember or forget information over long sequences. This architecture makes LSTMs particularly well-suited for time series forecasting, especially when dealing with complex, non-linear patterns and long-term dependencies. LSTMs can handle multivariate time series and can learn from historical data. However, they often require substantial training data to perform well, can be computationally intensive, and may be prone to overfitting if not properly regularized. Moreover, LSTM is a blackbox algorithm and can be challenging to interpret [33].

#### 3.1.5 Prophet

Prophet, developed by Meta (Facebook formally) in 2017 [34]. This algorithm offers a procedure for forecasting time series data based on an additive model that decomposes the time series into trend, seasonality, and holiday components. Prophet is designed to handle daily observations with at least one year of historical data and can accommodate missing values and outliers. The algorithm automatically detects changepoints in the trend and allows for user-specified changepoints. A key advantage of this algorithm is its ability to handle multiple seasonalities and incorporate domain knowledge through easily interpretable parameters. Prophet is particularly effective for forecasting tasks with strong seasonal effects (as well as those with several seasons of historical data). Yet, this algorithm may struggle with short-term forecasts or datasets with limited historical data. Additionally, while it is designed to be robust, it may not always capture complex, non-linear patterns effectively [35].

#### 3.1.6 Simple Moving Average (SMA)

The Simple Moving Average (SMA) is a basic and widely used time series forecasting method. The origin of SMA can be traced back to the early days of technical and inventory analysis [36]. This method calculates the arithmetic mean of a set of values over a specific number of time periods and is often used to smooth out short-term fluctuations and highlight longer-term trends or cycles, and can be effective for short-term forecasting in stable time series with minimal trend or seasonality. However, SMA has several limitations. This method can produce lags behind the most recent data points and may miss sudden changes or turning points. SMA also gives equal weight to all observations within the moving window, which may not be ideal if more recent observations are believed to be more relevant. Despite such limitations, SMA remains a useful tool for forecasting methods [37].

#### 3.1.7 Theta Method

The Theta algorithm was proposed by Assimakopoulos and Nikolopoulos in 2000 [38]. This algorithm decomposes the time series into two "theta lines." The first line represents the long-term trend, and the other captures short-term behavior. These lines are then extrapolated separately and combined to produce the final forecast. The Theta method is praised for its simplicity and effectiveness, especially for seasonal time series. The Theta algorithm often performs well without requiring extensive parameter tuning, making it accessible for practitioners. However, the method assumes that the time series can be well-represented via decomposition into two lines (which may not always hold true for complex, non-linear time series). Moreover, it may struggle with abrupt changes or structural breaks in the data [39].

#### 3.1.8 Vector Autoregression (VAR)

Vector Autoregression (VAR), introduced by Sims in 1980 [40], is a multivariate forecasting technique that extends the univariate autoregressive model to capture the linear interdependencies among multiple time series. In VAR, each variable is a linear function of past lags of itself and past lags of the other variables. This methodology makes VAR particularly useful for understanding the relationships between multiple related time series and generating forecasts for these interactions. VAR models are widely used in econometrics and financial time series forecasting as they can model feedback effects and provide insights into the dynamics between variables through tools like impulse response functions [41]. However, VAR models assume linear relationships between variables and can become overparameterized when dealing with many variables or long lag structures. This could potentially lead to poor forecasts [42].

*3.2 ML algorithms adapted for time series forecasting*

3.2.1 Gaussian Process Regression (GPR)

Gaussian Process Regression (GPR) is a non-parametric probabilistic approach to regression and time series forecasting that is rooted in Bayesian statistics. This algorithm was formalized for ML applications by Rasmussen and Williams [43]. The method models the target variable as a Gaussian process, assuming that any finite collection of data points has a multivariate Gaussian distribution. GPR is particularly valuable in time series forecasting for its ability to provide uncertainty estimates along with predictions [44]. It can capture complex, non-linear relationships in the data and handles missing values naturally [45]. The flexible method can incorporate various trends and seasonal patterns by choosing kernel functions. However, GPR can be computationally intensive for large datasets due to the need to invert large covariance matrices, and its performance depends on the choice of kernel function, which may require domain expertise or extensive selection procedures [46].

3.2.2 Gradient Boosting and CatBoost

Gradient Boosting stems from a family of ensemble learning techniques, and CatBoost was recently developed by Yandex [47]. These methods work by building a series of weak learners (typically decision trees) sequentially, with each learner trying to correct the errors of its predecessors. In time series contexts, gradient boosting methods can capture complex, non-linear relationships and handle multiple input variables [48]. CatBoost, in particular, is designed to reduce overfitting and handle categorical variables efficiently, which can be beneficial in forecasting scenarios. However, gradient boosting algorithms do not inherently account for the temporal ordering of data, requiring careful feature engineering to incorporate time-based information. As such, they may also struggle with capturing long-term dependencies without extensive lag features [49].

3.2.3 K-Nearest Neighbors (KNN)

The K-Nearest Neighbors algorithm is often deployed in regression and classification tasks and can be adapted for time series forecasting [50]. The algorithm is non-parametric and can capture non-linear patterns in the data. In the context of time series forecasting, KNN finds historical periods most similar to the current state and uses their subsequent values to make predictions. KNN can be particularly effective when the time series exhibits recurring patterns or when there are strong analogies between past and future behavior. However, this algorithm's performance can degrade with high-dimensional data and long-term forecasts (especially with the lack of strong trends). The choice of distance metric and the number of neighbors (k) can significantly impact forecast accuracy [51].

3.2.4 Neural Networks

Neural networks, encompassing various architectures beyond LSTM, have become increasingly popular for time series forecasting [52]. Neural networks can capture complex, non-linear relationships in time series data and are capable of handling multiple input variables. The flexibility of their design allows practitioners to tailor architectures to specific forecasting problems. However, neural networks are blackboxes that often require large amounts of training data to perform well and can be prone to overfitting if not properly regularized. Additionally, the selection of appropriate network architecture and hyperparameters often requires significant expertise and computational resources [53].

3.2.5 Random Forest, Bagging, and XGBoost

Random Forest, Bagging, and XGBoost are ensemble learning methods. Bagging, short for Bootstrap Aggregating is a method to reduce variance in predictive models by creating multiple subsets of the original dataset through bootstrap sampling. This method trains a separate model on each subset and aggregates their predictions. Random Forest, introduced by Breiman in 2001 [54], is a specific implementation of bagging that constructs multiple decision trees and merges their predictions to improve accuracy and control overfitting. XGBoost, developed by Chen and Guestrin in 2016 [55], implements gradient boosted decision trees designed for speed and performance. All these algorithms can handle non-linear relationships and are capable of capturing complex patterns in time series data when properly engineered features are provided. Additionally, while they can handle multiple input variables, they may struggle with capturing long-term dependencies without extensive lag features [56,57].

3.2.6 Support Vector Regression (SVR)

Support Vector Regression is an extension of Support Vector Machines (SVM) that was developed by Vapnik et al. in the 1990s [58]. In time series forecasting, SVR works by mapping the input data into a high-dimensional feature space and finding a hyperplane that best fits the data while maintaining a specified tolerance margin. SVR is capable of capturing non-linear relationships through the use of kernel functions, making it suitable for complex time series patterns. It is particularly effective when dealing with high-dimensional data and can handle multiple input variables. SVR is less prone to overfitting compared to some other ML algorithms due to its structural risk minimization principle [59]. However, the performance of SVR can be sensitive to the tuning of kernels/hyperparameters and necessitates careful feature engineering to incorporate time-based information [60].

Table 1 A comparison among the examined algorithms in this study

| Algorithm | Family | Methodology & Logic | Typical Use Cases | Strengths/Advantages | Weaknesses/Disadvantages |
|---|---|---|---|---|---|
| ARIMA | Statistical | Linear, combines differencing with autoregression and moving average components. | Time series data without seasonal patterns. | Flexible, good for none/some seasonal data. | Assumes linearity and stationarity, may not be suitable for complex patterns. |
| SARIMA | Statistical | Extends ARIMA to include seasonal components. | Seasonal time series data. | Handles seasonality, well-understood. | Computationally intensive, linear assumptions can be overfitted with short time series. |
| ETS | Statistical | Exponential smoothing (decomposing) techniques with error, trend, and seasonal components. | Short-term forecasting, seasonal and non-seasonal data. | Easy to implement, good for data with trends and seasonality. | May overfit on noisy data and can struggle with abrupt changes. |
| Holt-Winters | Statistical | Triple exponential smoothing for data with trends and seasonality. | Seasonal time series data. | Simple to implement, effective for additive and multiplicative seasonal patterns. | Assumes additive effects, may not handle high-frequency data well and can struggles with irregular time series |
| Prophet | Statistical | Additive and decomposable model with trend, seasonality, and holidays. | Daily data with strong multiple seasonality patterns, missing data, and outliers. | Robust to missing data, handles outliers, automatically detects changepoints, and | Less effective for non-daily data or complex patterns, and may struggle with short-term forecasts. |
| | | | | incorporates domain knowledge easily. | |
| Theta | Statistical | Decomposes data into two 'theta lines' for different trend assumptions (e.g., long and short-term components). | Time series data with trends. | Simple, effective for data with a trend. | Less effective for seasonal or non-linear data. Offers limited flexibility for complex patterns. |
| Simple Moving Average | Statistical | Calculates average over a fixed window of past observations (n). | Smoothing noisy data, simple forecasts. | Simple to understand and implement. | Not adaptive, lags in response to real trend changes. |
| VAR | Statistical | Vector Autoregression, multivariate linear model relating different time series variables. | Multivariate time series data. | Captures relationships between multiple series, good for stationary series. | Requires all series to be stationary, high computational cost. |
| Croston's Method | Statistical | Separately forecasts non-zero demand sizes and intervals and adjusts for intermittent demand. | Forecasting intermittent demand. | Good for sparse or intermittent data. | May be biased, assumes demand pattern does not change. |
| LSTM | ML | A type of recurrent neural network that uses gates to control information flow. | Complex patterns, large datasets, non-linear relationships. | Good for capturing long dependencies, non-linear patterns. | Requires large datasets, computationally intensive. Blackbox nature limits interpretability. |
| Neural Networks | ML | Layers of interconnected neurons learning data features. | Complex nonlinear patterns, high-dimensional data. | Highly flexible, powerful for complex patterns and can handle non-linear relationships. | Requires large data and careful feature engineering, prone to overfitting, black box. |
| Gaussian Process Regression | ML | Non-parametric kernel-based probabilistic model. | Small to medium datasets, needing uncertainty estimation. | Provides uncertainty measures, flexible. | Computationally expensive, not for large data. |
| KNN | ML | Predicts based on similar historical patterns by using 'k' nearest points for prediction. | Small datasets, simple non-linear patterns. | Simple, non-parametric, and effective for non-linearities in small datasets. | Not scalable, sensitive to the choice of k and noisy data. |
| SVR | ML | Fits within a certain threshold and finds the optimal hyperplane in high-dimensional space. | Regression with clear margin of error. | Effective in high-dimensional space, robust to outliers. | Requires good parameter tuning, and can be computationally intensive for very large datasets. |
| Random Forest | ML | Ensemble of decision trees, averaging to improve prediction. | Various problems. | Robust, handles overfitting well, good for mixed data types, and can provide feature importance | Requires feature engineering for temporal aspects. |
| XGBoost | ML | Gradient boosting with decision trees optimized for speed and performance. | Various problems. | Fast, scalable, high performance, handles various data types. | Prone to overfitting if not tuned properly. |
| Gradient Boosting | ML | Sequential correction of predecessor's errors, using decision trees. | Various problems. | Reduces bias and variance, powerful. | Computationally intensive and prone to overfitting. |
| CatBoost | ML | Categorical Boosting. | Datasets with many categorical features. | Efficient with categorical data, less prone to overfitting. | Slightly slower compared to other boosting methods. |
| Bagging | ML | Bootstrap aggregating, reduces variance by averaging a set of parallel estimators. | Reducing variance in noisy data sets. | Reduces overfitting, robust to noisy data. | Can be less effective on biased models, high memory consumption. |

## 4.0 Description of benchmarking experiments, metrics, and datasets

This benchmarking analysis involves a set of 25 synthetic and 25 real timeseries. This analysis was run and evaluated in a Python 3.10.5 environment using an Intel(R) Core(TM) i7-9700F CPU @ 3.00GHz and an installed RAM of 32.0GB. All algorithms were run in default settings to allow fairness and ensure reproducibility, and the performance of each algorithm was evaluated through several metrics, as discussed below and listed in Table 2. These metrics, along with the selected sizes of datasets, followed the recommendations of [7,61].

We utilize four primary metrics that can be classified under global/general and specific/internal metrics. The global metrics are suitable for broad comparisons and evaluations across multiple datasets and models (e.g., Mean Absolute Scaled Error (MASE) and Dynamic Time Warping (DTW)). On the other hand, internal metrics are used to provide detailed insights into particular aspects of model performance (such as mean absolute deviation (MAD) and direction accuracy (DA)). It is worth noting that other metrics (i.e., Root Mean Square Error (RMSE) and the Mean Absolute Error (MAE)) were not used herein due to their inherent limitations and vulnerabilities with respect to time series analysis, as pointed out by [62,63].

The MASE is a relative measure of forecast accuracy that scales the forecast error by the in-sample mean absolute error. MASE is scale-independent and can be used to compare forecast accuracy across different time series [64,65]. DTW measures the similarity between two time series by finding an optimal alignment between them. Unlike simple distance measures, DTW can handle time shifts and distortions by allowing flexible matching of time indices [66]. The MAD measures the average absolute error between the actual and forecasted values to clearly indicate the average magnitude of forecast errors [67]. The DA measures how well the model predicts the direction of the time series movement. This metric evaluates whether the forecast correctly predicts the increase or decrease in the actual values from one time point to the next.

Table 2 List of performance metrics.

| Type | Metric | Formula |
|---|---|---|
| Specific/Internal Metrics | Mean Absolute Deviation (MAD) | $MAD = \frac{1}{n}\sum_{t=1}^{n}\lvert y_t - \hat{y}_t\rvert$<br>Where:<br>• $y_t$ is the actual value at time t.<br>• $y\hat{}_t$ is the forecasted value at time t.<br>n is the total number of observations. |
| | Direction Accuracy (DA) | $DA = \frac{1}{n-1}\sum_{t=2}^{n} \text{II}\left((y_t - y_{t-1})(\hat{y}_t - \hat{y}_{t-1}) > 0\right)$<br>Where:<br>• $y_t$ is the actual value at time t.<br>• $y\hat{}_t$ is the forecasted value at time t.<br>• II is the indicator function that equals 1 if the condition inside is true and 0 otherwise.<br>• n is the total number of observations. |
| Global/External Metrics | Mean Absolute Scaled Error (MASE) | $MASE = \frac{\frac{1}{n}\sum_{t=1}^{n}\lvert y_t - \hat{y}_t\rvert}{\frac{1}{n}\sum_{t=1}^{n}\lvert y_t - y_{t-1}\rvert}$<br>Where:<br>• $y_t$ is the actual value at time t.<br>• $y\hat{}_t$ is the forecasted value at time t.<br>• n is the total number of observations. |
| | Dynamic Time Warping (DTW) | $DTW(A,B) = min\sqrt{\sum_{i=2}^{n}(a_i - b_i)^2}$<br>Where:<br>• $A=(a_1,a_2,\ldots,a_n)$ and $B = (b_1,b_2,\ldots,b_m)$ are two sequences of length n and m respectively.<br>• i' is the optimal alignment index of b corresponding to $a_i$. |

*4.1 Synthetic datasets*

Twenty five synthetic timeseries of various scenarios were generated and examined by all algorithms (see Fig. 1 and Table 3). These timeseries were generated via the `generate_time_series` function, which allows researchers to generate controlled datasets that can be used to benchmark and evaluate the performance of time series forecasting models. This particular function accepts a specific mathematical function and the number of data points (`n_points`) as input parameters. Then, this function generates a sequence of equally spaced time points over a specified range (`t_max`). The chosen function is applied to these time points to produce the corresponding time series data. Gaussian noise is added to simulate real-world conditions where data often includes random variations. The `generate_time_series` function starts by creating an array of time points using `numpy.linspace`, which ensures an even distribution of points between 0 and the specified `t_max`. This array of time points, `t`, is then passed to the provided time series function (func), which applies the mathematical transformation and returns the resulting data series.

Table 3 Parameters used in the synthetic timeseries

| Function | Description | Mathematical Expression | Noise Level | Characteristics |
|---|---|---|---|---|
| Linear trend | Linear increase with Gaussian noise | $0.5t+\epsilon$ | $\sigma=0.1$ | Simple trend |
| Quadratic trend | Quadratic increase with Gaussian noise | $0.05t^2+\epsilon$ | $\sigma=0.1$ | Parabolic trend |
| Exponential growth | Exponential increase with Gaussian noise | $e^{0.1t}+\epsilon$ | $\sigma=0.1$ | Exponential trend |
| Sine wave (seasonal) | Periodic sine wave with Gaussian noise | $\sin(2\pi t)+\epsilon$ | $\sigma=0.1$ | Seasonal, periodic pattern |
| Cosine wave with linear trend | Cosine wave superimposed on a linear trend with noise | $\cos(2\pi t)+0.1t+\epsilon$ | $\sigma=0.1$ | Trend and seasonality combination |
| Composite of sine waves | Multiple sine waves combined with Gaussian noise | $\sin(2\pi t)+0.5\sin(4\pi t)+\epsilon$ | $\sigma=0.1$ | Multiple seasonalities |
| Logistic growth | Sigmoidal growth with Gaussian noise | $1/1+e-t+5+\epsilon$ | $\sigma=0.05$ | Non-linear growth |
| Damped oscillation | Exponentially damped sine wave with noise | $e^{-0.1t}\sin(2\pi t)+\epsilon$ | $\sigma=0.05$ | Oscillatory effect |
| Step function | Discrete steps with Gaussian noise | $step(t)+\epsilon$ | $\sigma=0.1$ | Abrupt changes |
| Sawtooth wave | Linear periodic rise with a drop and Gaussian noise | $(t\%1)+\epsilon$ | $\sigma=0.05$ | Sharp transitions |
| Square wave | Alternating high and low values with Gaussian noise | $sign(\sin(2\pi t))+\epsilon$ | $\sigma=0.1$ | Discrete, binary states |
| Exponential decay | Exponential decrease with Gaussian noise | $e^{-0.2t}+\epsilon$ | $\sigma=0.05$ | Decay trend |
| Logarithmic growth | Logarithmic increase with Gaussian noise | $\log(t+1)+\epsilon$ | $\sigma=0.1$ | Logarithmic trend |
| Composite trend, seasonal, and noise | Combination of quadratic trend, sine wave, and noise | $0.01t^2+\sin(2\pi t)+0.5\epsilon$ | $\sigma=1$ | Complex pattern |
| Autocorrelated process (AR(1)) | Autoregressive process with Gaussian noise | $0.8y_{t-1}+\epsilon$ | $\sigma=0.5$ | Dependency on previous values |
| Polynomial trend (cubic) | Cubic polynomial trend with Gaussian noise | $0.01t^3-0.1t^2+0.5t+\epsilon$ | $\sigma=0.1$ | Higher-order polynomial trend |
| Sigmoid function | Sigmoidal growth with Gaussian noise | $1/1+e^{-t+5}+\epsilon$ | $\sigma=0.05$ | Non-linear growth |
| Impulse response | Exponentially decaying sinusoidal impulse with noise | $e^{-t}\sin(2\pi t)+\epsilon$ | $\sigma=0.05$ | Impulse-like behavior |
| Cyclical pattern with trend | Sine wave with linear trend and Gaussian noise | $\sin(2\pi t/5)+0.05t+\epsilon$ | $\sigma=0.1$ | Cyclic and trending |
| Composite of exponential growth and seasonal pattern | Exponential growth with superimposed sine wave and noise | $e^{0.05t}+0.5\sin(2\pi t)+\epsilon$ | $\sigma=0.1$ | Complex trend and seasonality |
| Piecewise linear function | Linear segments with Gaussian noise | $piecewise(t)+\epsilon$ | $\sigma=0.1$ | Segmented linear behavior |
| Brownian motion (random walk) | Cumulative sum of Gaussian noise | $\sum\epsilon$ | $\sigma=0.1$ | Stochastic, random walk |
| Composite of multiple trends | Combination of quadratic, sinusoidal, and exponential trends with noise | $0.01t^2+0.1\sin(2\pi t)+0.05e^{0.1t}+\epsilon$ | $\sigma=0.1$ | Multiple trend components |
| Chaotic logistic map | Logistic map function with chaos | 3.9t(1−t)+ϵ | σ=0.1 | Chaotic behavior |
| GARCH-like volatility clustering | Gaussian noise with time-varying volatility | N[(0, 0.1 + 0.9 abs($y_{t-1}$))] | (0, 0.1 + 0.9 abs($y_{t-1}$) | Volatility clustering |

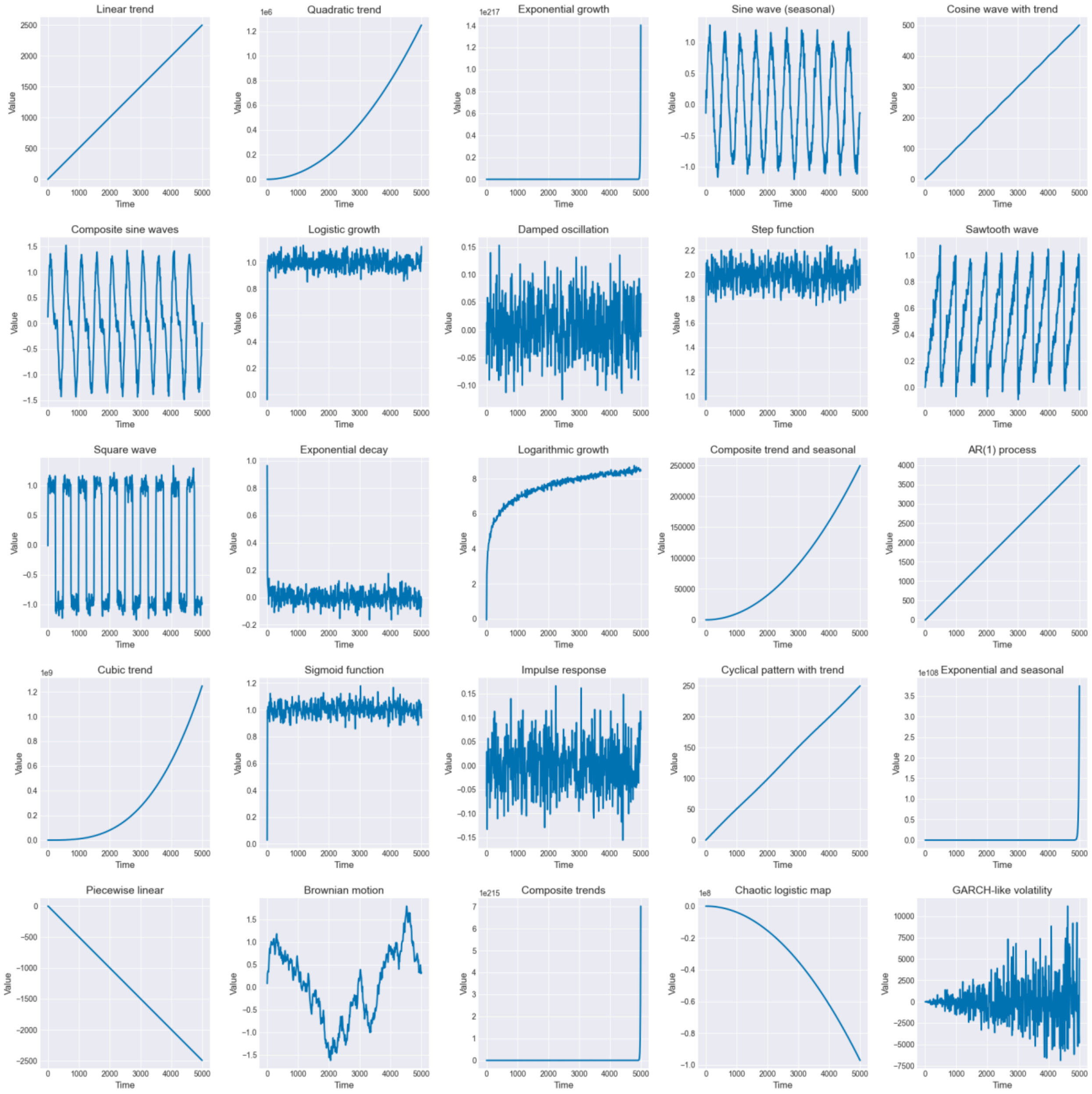

Fig. 1 Visualization of the synthetic datasets

The outcome of the benchmarking analysis on the synthetic datasets is listed in Table 4. As one can see, this table showcases three different ranking methods (namely, based on the average ranking, normalized ranking, and wins). All ranking systems used the abovementioned DA, DTW, MASE, and MAD metrics, wherein lower values indicate better performance (rank 1 is best), except for the DA metric, where higher values indicate better performance (rank 1 is best).

The first ranking system represents averages across all datasets for each algorithm. Each metric was ranked individually, and the average of these ranks determined the Average Rank for each algorithm. The Final Rank was then assigned based on the Average Rank values. Then, the second ranking system involves averaging the original metric values across all datasets for each algorithm. These averages were then normalized to a 0-1 scale for each metric. The Average Normalized Score is then computed as the mean of all normalized metric scores for each algorithm, and the Final Rank is derived from these average normalized scores to allow for a fair comparison across different metrics and algorithms. Finally, the third ranking system adopted the ranking by Wins method. In this system, metric values were averaged across all datasets for each algorithm. Each metric was ranked individually. The rank columns display the rank for each metric, and the Average Rank reflects the mean of these ranks for each algorithm. The Final Rank was determined based on the Average Rank values. This ranking method aimed to balance performance across all metrics, emphasizing how frequently each algorithm performed best in each metric.

Table 4 presents a collective view of the algorithmic performance across the different datasets and systems used. As one can see, SPINEX ranks 5$^{th}$ under the first ranking system and 1$^{st}$ under the other two ranking systems. Despite not being the top-ranked algorithm in the first ranking system, SPINEX consistently performed well across different metrics. This performance suggests a well-rounded response. When comparing SPINEX to other algorithms such as SARIMA, Prophet, Holt-Winters, and Theta, it is evident that SPINEX stands out regarding consistent performance and robustness. Similarly, Prophet and Theta showed competitive performance but could not match SPINEX's consistency across all ranking systems.

Table 4 Ranking results on real data

| Algorithm | Direction Accuracy | DTW | MASE | MAD | Direction Accuracy (rank) | DTW (rank) | MASE (rank) | MAD (rank) | Average (rank) | Final (rank) |
|---|---|---|---|---|---|---|---|---|---|---|
| Based on average rankings | | | | | | | | | | |
| SARIMA | 0.578 | 15.353 | 2625.313 | 0.116 | 3.570 | 5.830 | 7.890 | 9.250 | 6.640 | 1 |
| Prophet | 0.546 | 2.058 | 58.910 | 0.095 | 3.870 | 6.210 | 7.430 | 9.030 | 6.640 | 2 |
| Holt-Winters | 0.572 | 15.419 | 46.643 | 0.291 | 3.360 | 6.400 | 7.980 | 8.910 | 6.660 | 3 |
| Theta | 0.550 | 2.451 | 82.481 | 0.100 | 3.600 | 7.080 | 7.520 | 9.660 | 6.960 | 4 |
| SPINEX | 0.602 | 1.956 | 45.676 | 0.075 | 3.230 | 6.840 | 10.300 | 8.950 | 7.330 | 5 |
| ARIMA | 0.264 | 2.544 | 84.369 | 0.097 | 7.190 | 8.800 | 7.900 | 6.980 | 7.720 | 6 |
| Croston | 0.000 | 2.602 | 87.962 | 0.101 | 10.370 | 9.130 | 7.850 | 5.360 | 8.180 | 7 |
| ETS | 0.000 | 2.603 | 87.962 | 0.101 | 10.370 | 9.360 | 7.980 | 5.350 | 8.260 | 8 |
| LSTM | 0.502 | 3.756 | 115.265 | 0.090 | 4.390 | 9.060 | 9.500 | 10.140 | 8.270 | 9 |
| Random Forest | 0.000 | 2.698 | 88.205 | 0.101 | 10.370 | 10.120 | 8.990 | 5.290 | 8.690 | 10 |
| Bagging | 0.000 | 2.698 | 88.205 | 0.101 | 10.370 | 10.120 | 8.990 | 5.290 | 8.690 | 10 |
| Gradient Boosting | 0.000 | 2.689 | 88.144 | 0.101 | 10.370 | 10.210 | 8.950 | 5.470 | 8.750 | 12 |
| XGBoost | 0.000 | 2.695 | 88.281 | 0.101 | 10.370 | 10.560 | 9.310 | 5.340 | 8.890 | 13 |
| SMA | 0.000 | 2.669 | 88.733 | 0.101 | 10.370 | 10.530 | 9.640 | 5.400 | 8.980 | 14 |
| KNN | 0.000 | 2.669 | 88.733 | 0.101 | 10.370 | 10.530 | 9.590 | 5.450 | 8.990 | 15 |
| CatBoost | 0.000 | 2.666 | 89.402 | 0.101 | 10.370 | 10.990 | 9.820 | 5.330 | 9.130 | 16 |
| Neural Network | 0.518 | 6.952 | 194.656 | 0.110 | 3.660 | 13.880 | 14.080 | 11.850 | 10.870 | 17 |
| SVR | 0.166 | 4.262 | 157.608 | 0.115 | 7.750 | 13.230 | 14.220 | 11.280 | 11.620 | 18 |
| Gaussian Process | 0.255 | 7.354 | 144.260 | 0.097 | 6.840 | 15.660 | 16.440 | 9.300 | 12.060 | 19 |
| Based on normalized rankings | | | | | | | | | | |
| SPINEX | 0.602 | 1.956 | 45.676 | 0.075 | 0.000 | 0.000 | 0.000 | 0.000 | 0.000 | 1 |
| Prophet | 0.546 | 2.058 | 58.910 | 0.095 | 0.094 | 0.008 | 0.005 | 0.093 | 0.050 | 2 |
| Theta | 0.550 | 2.451 | 82.481 | 0.100 | 0.087 | 0.037 | 0.014 | 0.115 | 0.063 | 3 |
| LSTM | 0.502 | 3.756 | 115.265 | 0.090 | 0.167 | 0.134 | 0.027 | 0.070 | 0.100 | 4 |
| ARIMA | 0.264 | 2.544 | 84.369 | 0.097 | 0.562 | 0.044 | 0.015 | 0.103 | 0.181 | 5 |
| Neural Network | 0.518 | 6.952 | 194.656 | 0.110 | 0.139 | 0.371 | 0.058 | 0.164 | 0.183 | 6 |
| Gaussian Process | 0.255 | 7.354 | 144.260 | 0.097 | 0.577 | 0.401 | 0.038 | 0.103 | 0.280 | 7 |

| | | | | | | | | | |
|---|---|---|---|---|---|---|---|---|---|
| SVR | 0.166 | 4.262 | 157.608 | 0.115 | 0.725 | 0.171 | 0.043 | 0.184 | 0.281 | 8 |
| Croston | 0.000 | 2.602 | 87.962 | 0.101 | 1.000 | 0.048 | 0.016 | 0.119 | 0.296 | 9 |
| ETS | 0.000 | 2.603 | 87.962 | 0.101 | 1.000 | 0.048 | 0.016 | 0.119 | 0.296 | 10 |
| KNN | 0.000 | 2.669 | 88.733 | 0.101 | 1.000 | 0.053 | 0.017 | 0.119 | 0.297 | 11 |
| SMA | 0.000 | 2.669 | 88.733 | 0.101 | 1.000 | 0.053 | 0.017 | 0.119 | 0.297 | 11 |
| CatBoost | 0.000 | 2.666 | 89.402 | 0.101 | 1.000 | 0.053 | 0.017 | 0.119 | 0.297 | 13 |
| Gradient Boosting | 0.000 | 2.689 | 88.144 | 0.101 | 1.000 | 0.054 | 0.016 | 0.119 | 0.297 | 14 |
| XGBoost | 0.000 | 2.695 | 88.281 | 0.101 | 1.000 | 0.055 | 0.017 | 0.119 | 0.298 | 15 |
| Random Forest | 0.000 | 2.698 | 88.205 | 0.101 | 1.000 | 0.055 | 0.016 | 0.119 | 0.298 | 16 |
| Bagging | 0.000 | 2.698 | 88.205 | 0.101 | 1.000 | 0.055 | 0.016 | 0.119 | 0.298 | 16 |
| Holt-Winters | 0.572 | 15.419 | 46.643 | 0.291 | 0.051 | 1.000 | 0.000 | 1.000 | 0.513 | 18 |
| SARIMA | 0.578 | 15.353 | 2625.313 | 0.116 | 0.040 | 0.995 | 1.000 | 0.192 | 0.557 | 19 |
| Based on wins | | | | | | | | | | |
| SPINEX | 0.602 | 1.956 | 45.676 | 0.075 | 1 | 1 | 1 | 1 | 1 | 1 |
| Prophet | 0.546 | 2.058 | 58.910 | 0.095 | 5 | 2 | 3 | 3 | 3.25 | 2 |
| Theta | 0.550 | 2.451 | 82.481 | 0.100 | 4 | 3 | 4 | 6 | 4.25 | 3 |
| ARIMA | 0.264 | 2.544 | 84.369 | 0.097 | 8 | 4 | 5 | 4 | 5.25 | 4 |
| Croston | 0.000 | 2.602 | 87.962 | 0.101 | 11 | 5 | 6 | 7 | 7.25 | 5 |
| ETS | 0.000 | 2.603 | 87.962 | 0.101 | 11 | 6 | 7 | 7 | 7.75 | 6 |
| Gradient Boosting | 0.000 | 2.689 | 88.144 | 0.101 | 11 | 10 | 8 | 7 | 9 | 7 |
| SMA | 0.000 | 2.669 | 88.733 | 0.101 | 11 | 8 | 12 | 7 | 9.5 | 8 |
| LSTM | 0.502 | 3.756 | 115.265 | 0.090 | 7 | 14 | 15 | 2 | 9.5 | 8 |
| KNN | 0.000 | 2.669 | 88.733 | 0.101 | 11 | 8 | 12 | 7 | 9.5 | 8 |
| Random Forest | 0.000 | 2.698 | 88.205 | 0.101 | 11 | 12 | 9 | 7 | 9.75 | 11 |
| CatBoost | 0.000 | 2.666 | 89.402 | 0.101 | 11 | 7 | 14 | 7 | 9.75 | 11 |
| Bagging | 0.000 | 2.698 | 88.205 | 0.101 | 11 | 12 | 9 | 7 | 9.75 | 11 |
| XGBoost | 0.000 | 2.695 | 88.281 | 0.101 | 11 | 11 | 11 | 7 | 10 | 14 |
| Holt-Winters | 0.572 | 15.419 | 46.643 | 0.291 | 3 | 19 | 2 | 19 | 10.75 | 15 |
| Gaussian Process | 0.255 | 7.354 | 144.260 | 0.097 | 9 | 17 | 16 | 5 | 11.75 | 16 |
| Neural Network | 0.518 | 6.952 | 194.656 | 0.110 | 6 | 16 | 18 | 16 | 14 | 17 |
| SARIMA | 0.578 | 15.353 | 2625.313 | 0.116 | 2 | 18 | 19 | 18 | 14.25 | 18 |
| SVR | 0.166 | 4.262 | 157.608 | 0.115 | 10 | 15 | 17 | 17 | 14.75 | 19 |

Figure 2 shows a more detailed examination of the performance of all algorithm algorithms evaluated across different settings and metrics. This evaluation was conducted for two different parameters: maximum time ($t_{max}$) and number of sequence points ($n_{points}$).

In terms of DA, which measures how well the model predicts the direction of the time series movement, SPINEX maintained strong performance across both settings. The graphs indicate that SPINEX's performance remained relatively stable and high compared to other algorithms as the sub-settings increased. Such stability can be crucial for applications requiring reliable directional predictions. More specifically, in the $t_{max}$ graph, SPINEX showed a slight improvement with higher sub-settings, indicating its adaptability to longer forecasting horizons. Similarly, in the $n_{points}$ graph, SPINEX outperformed most other algorithms, demonstrating its effectiveness in handling varying data point quantities.

For the DTW metric, which measures the alignment between predicted and actual time series, SPINEX also performed well across different settings. In both graphs, SPINEX maintained lower DTW values, indicating closer alignment and better predictive accuracy. The $t_{max}$ graph shows that SPINEX's DTW values remained relatively stable, suggesting its robustness to changes in the forecast length. The $n_{points}$ graph further highlights SPINEX's capability to handle datasets with varying numbers of points without significant loss in accuracy. Furthermore, SPINEX maintained lower MASE and MAD values compared to many other algorithms. This performance indicates that SPINEX can provide accurate forecasts. Figure 3 presents a visual example of two time series as predicted by SPINEX and other algorithms.

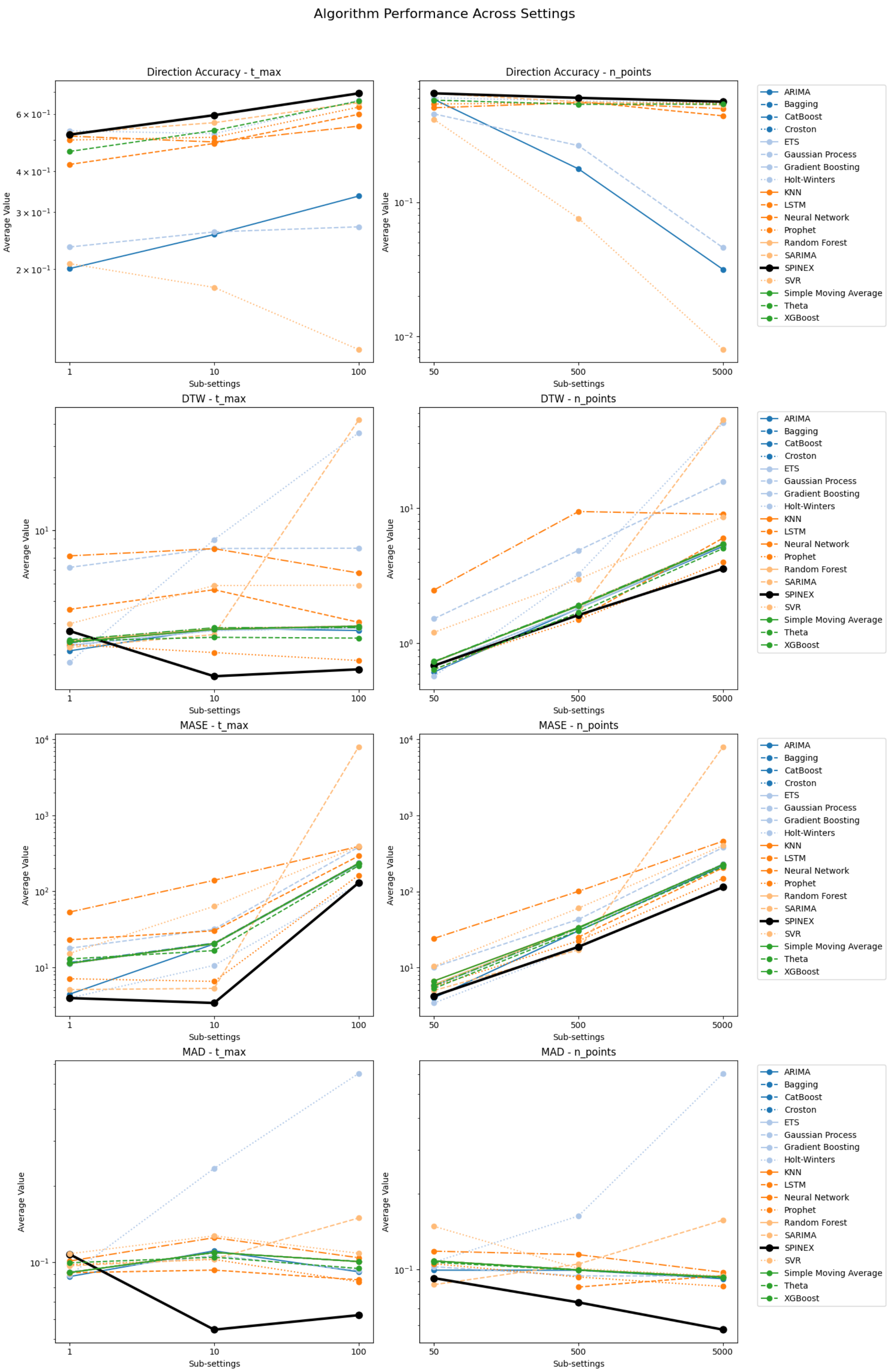

Figure 2 Individual rankings per algorithm for the internal and external metrics.

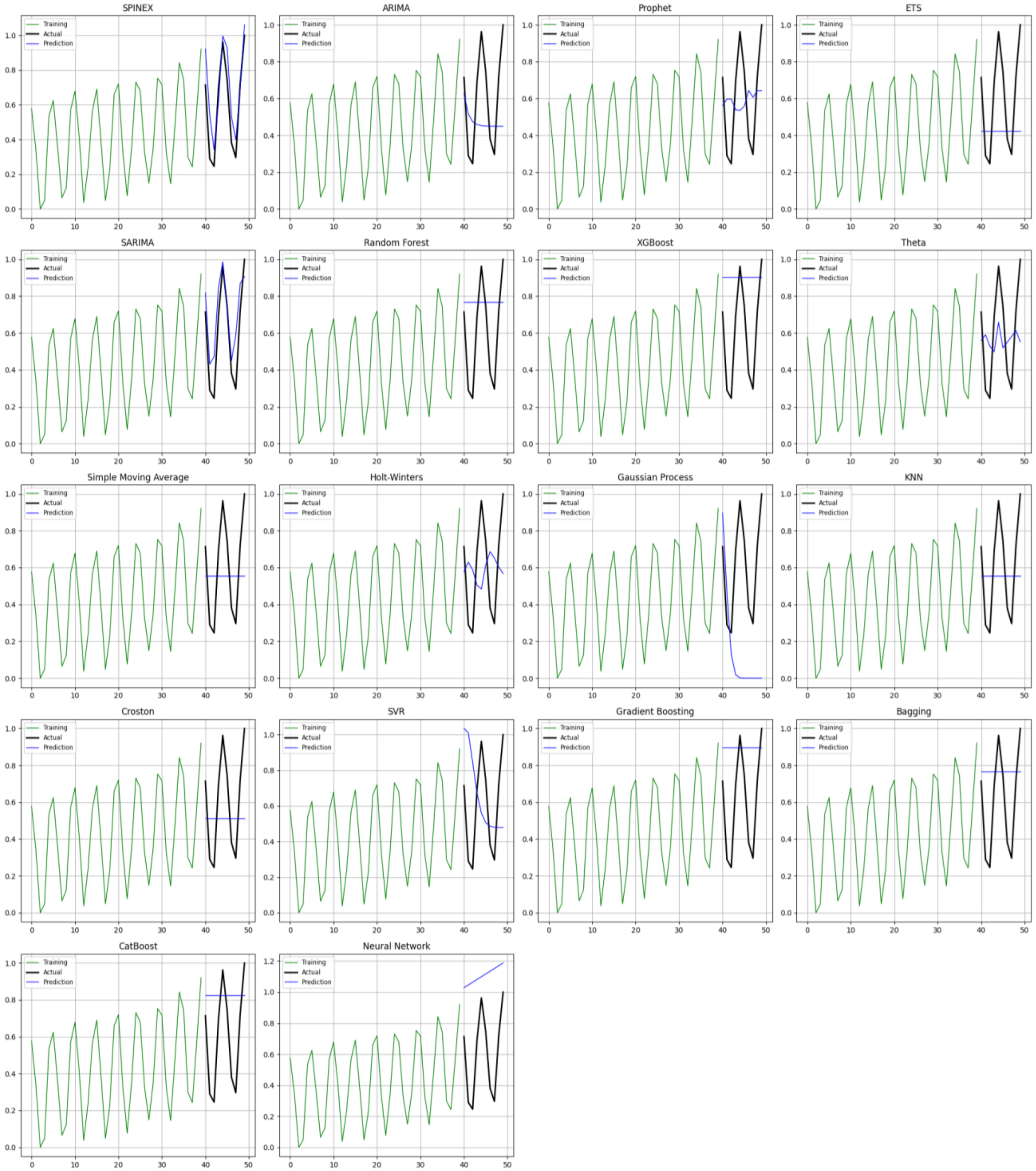

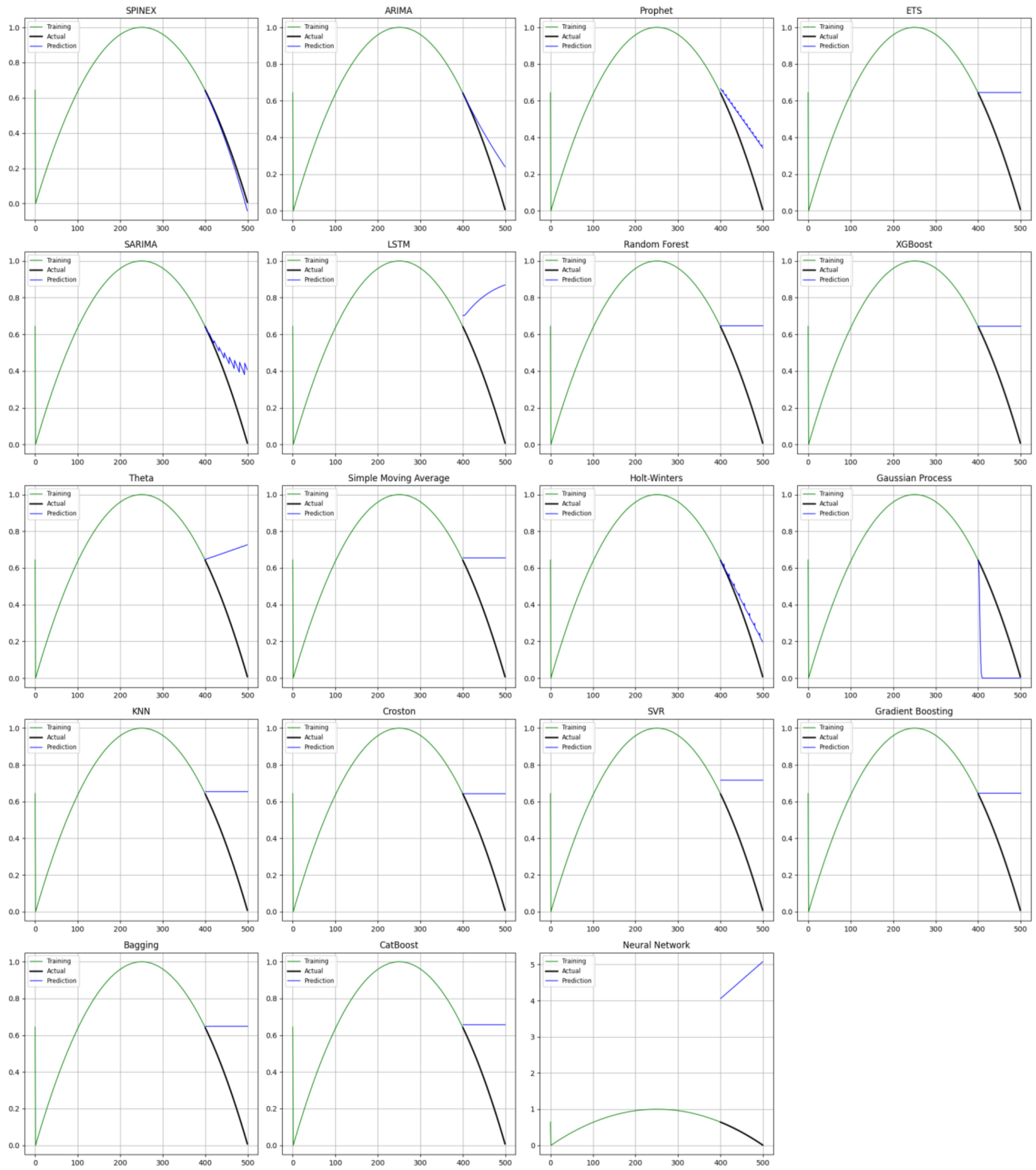

Fig. 3 Visualization of forecasting on Dataset no. 2 [$t_{max}$ = 10, $n_{points}$ = 50] (top) and Dataset no. 6 [$t_{max}$ = 1, $n_{points}$ = 500] (bottom)

*4.2 Real datasets*

Twenty four real datasets were used herein to further evaluate the performance of SPINEX against the other algorithms listed above. These datasets span univariate and multivariate scenarios(see Table 5 and Fig. 4). Additional details can be found in the cited sources.

Table 5 Real datasets used in the analysis

| Type | Dataset Name | Samples* | Features | References |
|---|---|---|---|---|
| Univariate | Airline Passengers | 144 | 2 | [68] |
| | Sunspots | 2820 | 2 | [69] |
| | Daily Female Births | 365 | 2 | [70] |
| | Yearly Water Usage | 79 | 2 | [71] |
| | Daily Minimum Temperatures | 3650 | 2 | [72] |
| | Monthly Car Sales | 108 | 2 | [73] |
| | Shampoo Sales Data | 36 | 2 | [74] |
| | Temperature Data | 3650 | 2 | [75] |
| | Monthly Writing Paper Sales | 147 | 2 | [76] |
| | Monthly Champagne Sales | 105 | 2 | [77] |
| | Monthly Robberies | 118 | 2 | [78] |
| | Electric Production | 397 | 2 | [79] |
| | Web Traffic Dataset | 550 | 2 | [80] |
| Multivariate | Stock and PM2.5 Prediction | 5650 | 10 | [81] |
| | Tata Global Forecasting | 2100 | 8 | [82] |
| | International Airline Passengers | 13391 | 6 | [83] |
| | Pollution Dataset | 43824 | 13 | [84] |
| | Daily Stock Prices | 52000 | 8 | [85] |
| | ETT-small | 17420 | 8 | [86] |
| | Jaipur Final Clean Data | 676 | 40 | [87] |
| | Aprocessed | 604802 | 17 | [88] |
| | Insurance | 1338 | 7 | [89] |
| | Indian Crime Data Analysis Forecasting I | 9840 | 33 | [90] |
| | Indian Crime Data Analysis Forecasting II | 295374 | 3 | [90] |

*Large datasets were stopped at 5000 data points, given the computational resources available during this study.

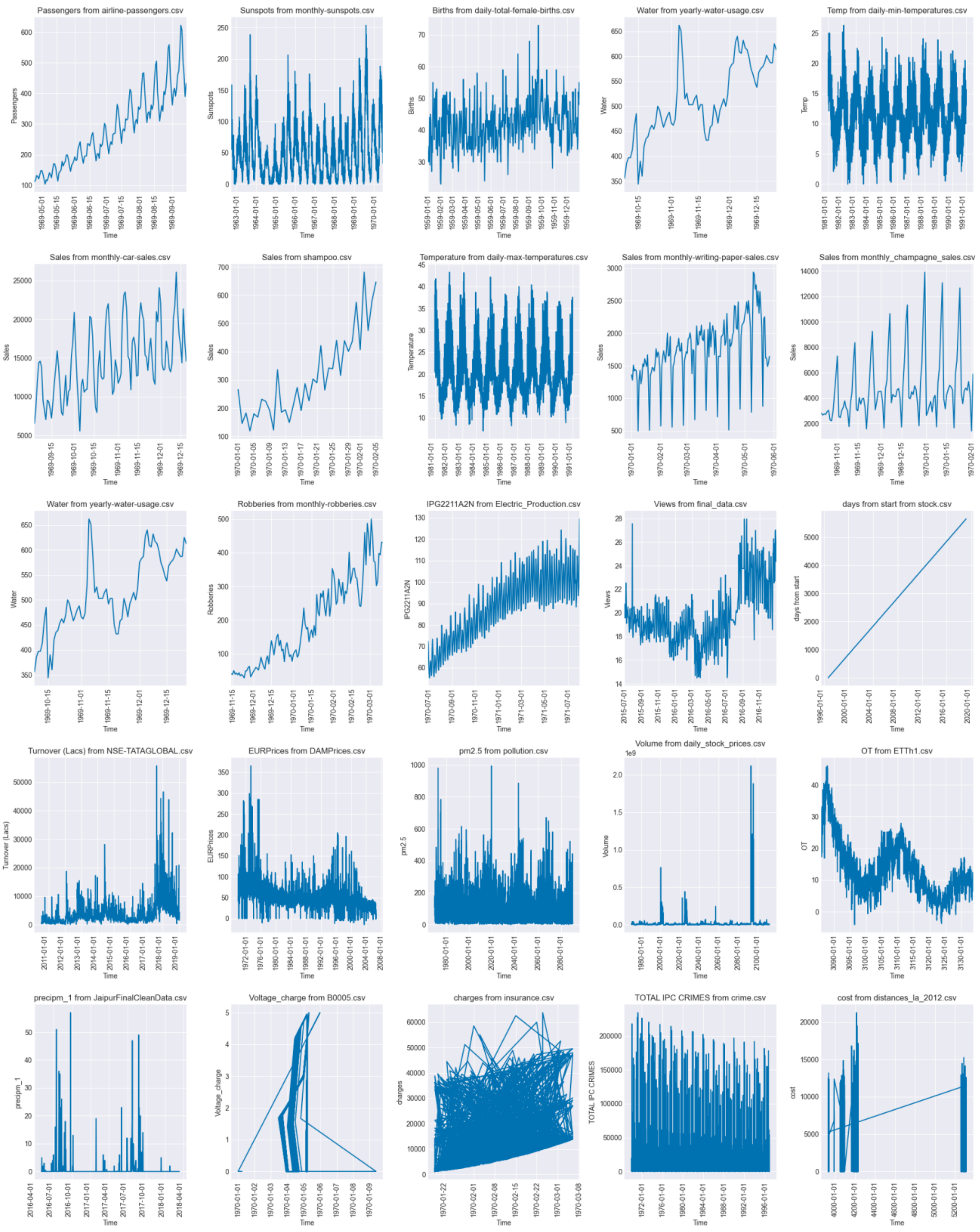

Fig. 4 Visualization of the real datasets

The benchmarking and ranking analysis results are examined similarly to the case of synthetic data by using the average, normalized, and win methods. These results are listed in Table 6. This table shows that SPINEX consistently ranks in the top two positions, along with the Holt-Winters algorithm.

Table 6 Ranking results on real data

| Algorithm | Direction Accuracy | DTW | MASE | MAD | Direction Accuracy (rank) | DTW (rank) | MASE (rank) | MAD (rank) | Overall rank | Final (rank) |
|---|---|---|---|---|---|---|---|---|---|---|
| Based on average rankings | | | | | | | | | | |
| Holt-Winters | 0.549 | 1.998 | 3.807 | 0.060 | 5.400 | 9.960 | 13.960 | 15.960 | 11.320 | 1 |
| SPINEX | 0.563 | 1.635 | 2.671 | 0.063 | 5.320 | 5.080 | 17.240 | 18.520 | 11.540 | 2 |
| SARIMA | 0.540 | 2.422 | 3.778 | 0.065 | 5.400 | 11.160 | 16.040 | 15.800 | 12.100 | 3 |
| Prophet | 0.464 | 2.100 | 6.896 | 0.071 | 8.440 | 8.680 | 13.560 | 19.640 | 12.580 | 4 |
| Theta | 0.491 | 2.299 | 13.739 | 0.073 | 6.760 | 13.400 | 18.040 | 13.160 | 12.840 | 5 |
| SMA | 0.000 | 2.472 | 23.792 | 0.078 | 20.040 | 19.320 | 14.840 | 10.600 | 16.200 | 6 |
| LSTM | 0.402 | 15.077 | 48.378 | 0.075 | 9.500 | 17.420 | 20.670 | 17.830 | 16.350 | 7 |
| KNN | 0.000 | 2.472 | 23.792 | 0.078 | 20.040 | 19.720 | 15.160 | 10.600 | 16.380 | 8 |
| XGBoost | 0.000 | 2.594 | 24.067 | 0.078 | 20.040 | 20.200 | 16.840 | 10.280 | 16.840 | 9 |
| Gradient Boosting | 0.000 | 2.567 | 24.043 | 0.078 | 20.040 | 20.120 | 16.680 | 10.760 | 16.900 | 10 |
| Random Forest | 0.000 | 2.597 | 23.894 | 0.078 | 20.040 | 20.920 | 16.680 | 11.080 | 17.180 | 11 |
| Bagging | 0.000 | 2.597 | 23.894 | 0.078 | 20.040 | 20.920 | 16.680 | 11.080 | 17.180 | 11 |
| CatBoost | 0.000 | 2.629 | 24.776 | 0.078 | 20.040 | 21.640 | 18.200 | 9.960 | 17.460 | 13 |
| Croston | 0.000 | 2.610 | 23.925 | 0.078 | 20.040 | 21.400 | 19.400 | 10.840 | 17.920 | 14 |
| ETS | 0.000 | 2.612 | 23.927 | 0.078 | 20.040 | 21.800 | 19.480 | 10.840 | 18.040 | 15 |
| ARIMA | 0.146 | 2.536 | 23.839 | 0.079 | 15.400 | 21.000 | 20.440 | 17.640 | 18.620 | 16 |
| SVR | 0.156 | 3.348 | 105.681 | 0.093 | 15.080 | 21.480 | 23.080 | 19.880 | 19.880 | 17 |
| Neural Network | 0.489 | 5.460 | 37.621 | 0.084 | 6.360 | 27.320 | 26.040 | 20.920 | 20.160 | 18 |
| Gaussian Process | 0.197 | 6.240 | 159.080 | 0.082 | 13.080 | 33.640 | 32.200 | 19.560 | 24.620 | 19 |
| Based on normalized rankings | | | | | | | | | | |
| Holt-Winters | 0.549 | 1.998 | 3.807 | 0.060 | 0.024 | 0.027 | 0.007 | 0.000 | 0.015 | 1 |
| SPINEX | 0.563 | 1.635 | 2.671 | 0.063 | 0.000 | 0.000 | 0.000 | 0.093 | 0.023 | 2 |
| SARIMA | 0.540 | 2.422 | 3.778 | 0.065 | 0.039 | 0.059 | 0.007 | 0.176 | 0.070 | 3 |
| Prophet | 0.464 | 2.100 | 6.896 | 0.071 | 0.176 | 0.035 | 0.027 | 0.333 | 0.143 | 4 |
| Theta | 0.491 | 2.299 | 13.739 | 0.073 | 0.126 | 0.049 | 0.071 | 0.400 | 0.161 | 5 |
| Neural Network | 0.489 | 5.460 | 37.621 | 0.084 | 0.130 | 0.285 | 0.223 | 0.730 | 0.342 | 6 |
| ARIMA | 0.146 | 2.536 | 23.839 | 0.079 | 0.740 | 0.067 | 0.135 | 0.580 | 0.381 | 7 |
| Simple | 0.000 | 2.472 | 23.792 | 0.078 | 1.000 | 0.062 | 0.135 | 0.546 | 0.436 | 8 |
| KNN | 0.000 | 2.472 | 23.792 | 0.078 | 1.000 | 0.062 | 0.135 | 0.546 | 0.436 | 9 |
| Gradient Boosting | 0.000 | 2.567 | 24.043 | 0.078 | 1.000 | 0.069 | 0.137 | 0.546 | 0.438 | 10 |
| Bagging | 0.000 | 2.597 | 23.894 | 0.078 | 1.000 | 0.072 | 0.136 | 0.546 | 0.438 | 11 |
| Random | 0.000 | 2.597 | 23.894 | 0.078 | 1.000 | 0.072 | 0.136 | 0.546 | 0.438 | 11 |
| XGBoost | 0.000 | 2.594 | 24.067 | 0.078 | 1.000 | 0.071 | 0.137 | 0.546 | 0.438 | 13 |
| Croston | 0.000 | 2.610 | 23.925 | 0.078 | 1.000 | 0.072 | 0.136 | 0.546 | 0.439 | 14 |
| ETS | 0.000 | 2.612 | 23.927 | 0.078 | 1.000 | 0.073 | 0.136 | 0.546 | 0.439 | 15 |
| CatBoost | 0.000 | 2.629 | 24.776 | 0.078 | 1.000 | 0.074 | 0.141 | 0.546 | 0.440 | 16 |
| LSTM | 0.402 | 15.077 | 48.378 | 0.075 | 0.286 | 1.000 | 0.292 | 0.456 | 0.508 | 17 |
| SVR | 0.156 | 3.348 | 105.681 | 0.093 | 0.722 | 0.127 | 0.659 | 1.000 | 0.627 | 18 |
| Gaussian Process | 0.197 | 6.240 | 159.080 | 0.082 | 0.650 | 0.343 | 1.000 | 0.672 | 0.666 | 19 |
| Based on wins | | | | | | | | | | |
| SPINEX | 0.563 | 1.635 | 2.671 | 0.063 | 1 | 1 | 1 | 2 | 1.25 | 1 |
| Holt-Winters | 0.549 | 1.998 | 3.807 | 0.060 | 2 | 2 | 3 | 1 | 2 | 2 |
| SARIMA | 0.540 | 2.422 | 3.778 | 0.065 | 3 | 5 | 2 | 3 | 3.25 | 3 |
| Prophet | 0.464 | 2.100 | 6.896 | 0.071 | 6 | 3 | 4 | 4 | 4.25 | 4 |
| Theta | 0.491 | 2.299 | 13.739 | 0.073 | 4 | 4 | 5 | 5 | 4.5 | 5 |
| SMA | 0.000 | 2.472 | 23.792 | 0.078 | 11 | 6 | 6 | 7 | 7.5 | 6 |
| KNN | 0.000 | 2.472 | 23.792 | 0.078 | 11 | 7 | 7 | 7 | 8 | 7 |
| ARIMA | 0.146 | 2.536 | 23.839 | 0.079 | 10 | 8 | 8 | 16 | 10.5 | 8 |
| XGBoost | 0.000 | 2.594 | 24.067 | 0.078 | 11 | 10 | 14 | 7 | 10.5 | 8 |
| Croston | 0.000 | 2.610 | 23.925 | 0.078 | 11 | 13 | 11 | 7 | 10.5 | 8 |
| ETS | 0.000 | 2.612 | 23.927 | 0.078 | 11 | 14 | 12 | 7 | 11 | 11 |
| Random Forest | 0.000 | 2.597 | 23.894 | 0.078 | 11 | 11 | 9 | 13 | 11 | 11 |
| Bagging | 0.000 | 2.597 | 23.894 | 0.078 | 11 | 11 | 9 | 13 | 11 | 11 |
| Gradient Boosting | 0.000 | 2.567 | 24.043 | 0.078 | 11 | 9 | 13 | 13 | 11.5 | 14 |
| CatBoost | 0.000 | 2.629 | 24.776 | 0.078 | 11 | 15 | 15 | 7 | 12 | 15 |
| LSTM | 0.402 | 15.077 | 48.378 | 0.075 | 7 | 19 | 17 | 6 | 12.25 | 16 |
| Neural Network | 0.489 | 5.460 | 37.621 | 0.084 | 5 | 17 | 16 | 18 | 14 | 17 |
| Gaussian Process | 0.197 | 6.240 | 159.080 | 0.082 | 8 | 18 | 19 | 17 | 15.5 | 18 |
| SVR | 0.156 | 3.348 | 105.681 | 0.093 | 9 | 16 | 18 | 19 | 15.5 | 18 |

Figure 5 illustrates the performance of various algorithms across different settings and metrics. Each of these metrics is evaluated across two settings: dataset length (short [datasets of less than 200 points] vs. long [datasets of more than 200 points]) and dataset type (univariate vs. multivariate). Overall, one can see the performance of SPINEX matches well with other algorithms and, in some cases, outperforms them.

For example, in the DA plots, SPINEX demonstrates a notable trend for short and long sequences and univariate and multivariate data. This suggests that SPINEX is proficient at predicting the correct direction. In the DTW metric, SPINEX exhibits a lower DTW value for short sequences, which indicates a higher similarity and better alignment of time series data than other algorithms. However, as the sequence length extends, SPINEX's DTW value increases, suggesting that its ability to maintain similarity diminishes slightly with longer sequences. This observation holds for the uni and multivariate datasets and other algorithms.

The MASE metric plots reveal that SPINEX performs consistently well across different lengths and types, with slightly better performance for short and univariate sequences. This trend continues for long sequences, where SPINEX remains competitive. It is worth noting that this algorithm maintains the lowest average MASE for long and multivariate sequences among the other algorithms. Finally, the MAD metric shows that SPINEX consistently achieves low values across both length and type dimensions.

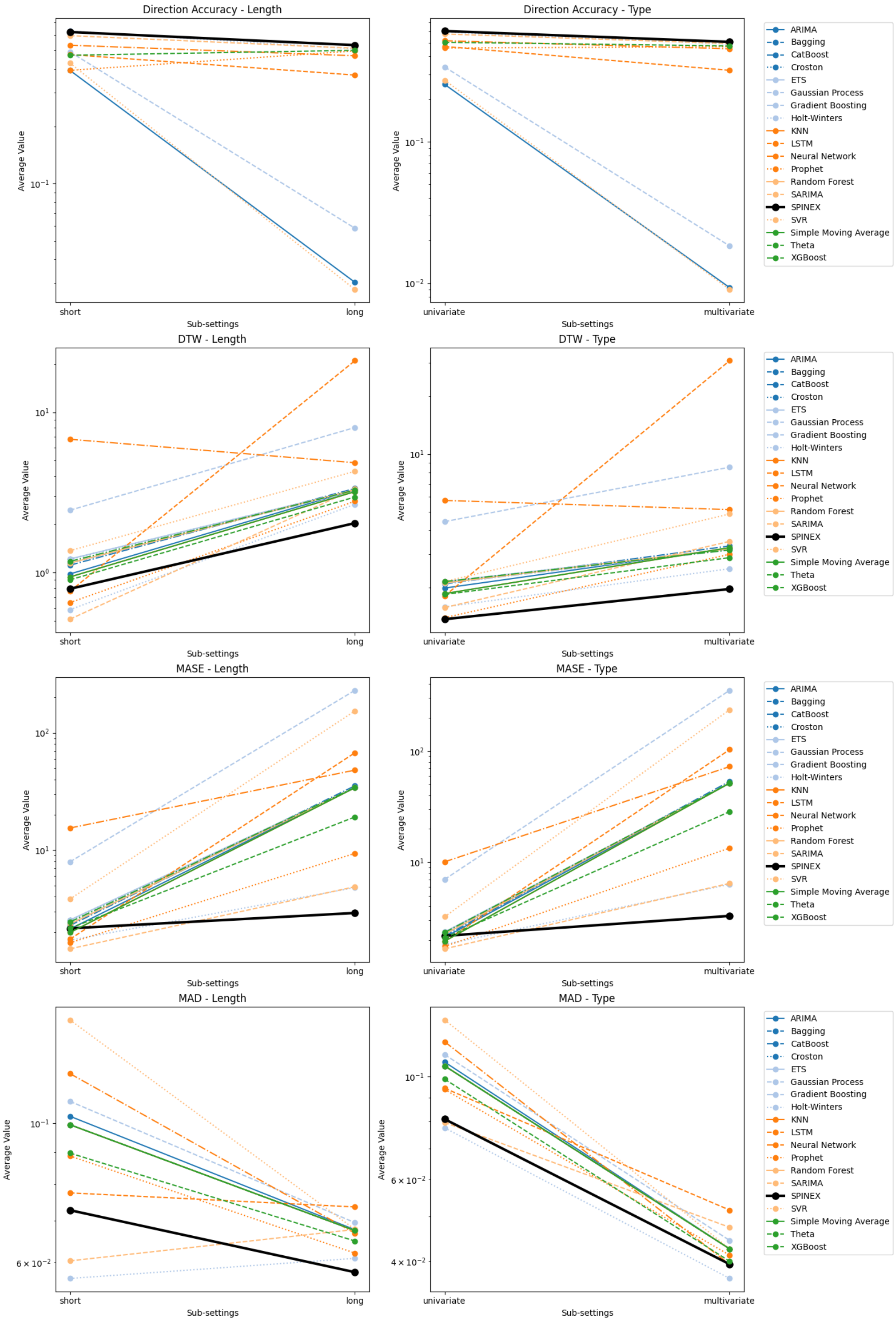

Figure 5 Further analysis in terms of dataset length and type

Figure 6 presents a sample of a visual representation of two time series as predicted by SPINEX and other algorithms. These two datasets represent those that fall under short and long time series. In both cases, it is clear that the forecasts by SPINEX are in good agreement with the actual series.

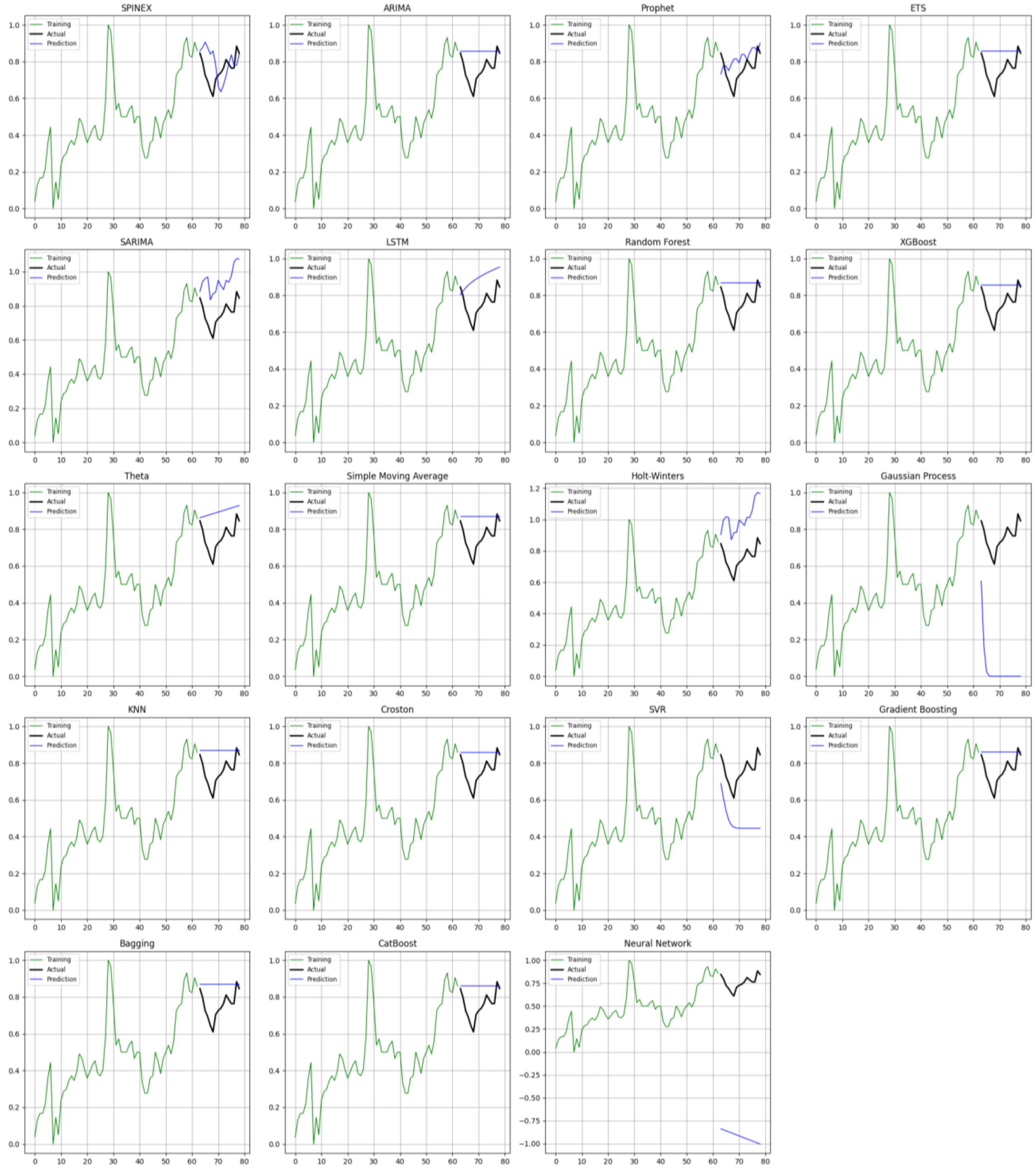

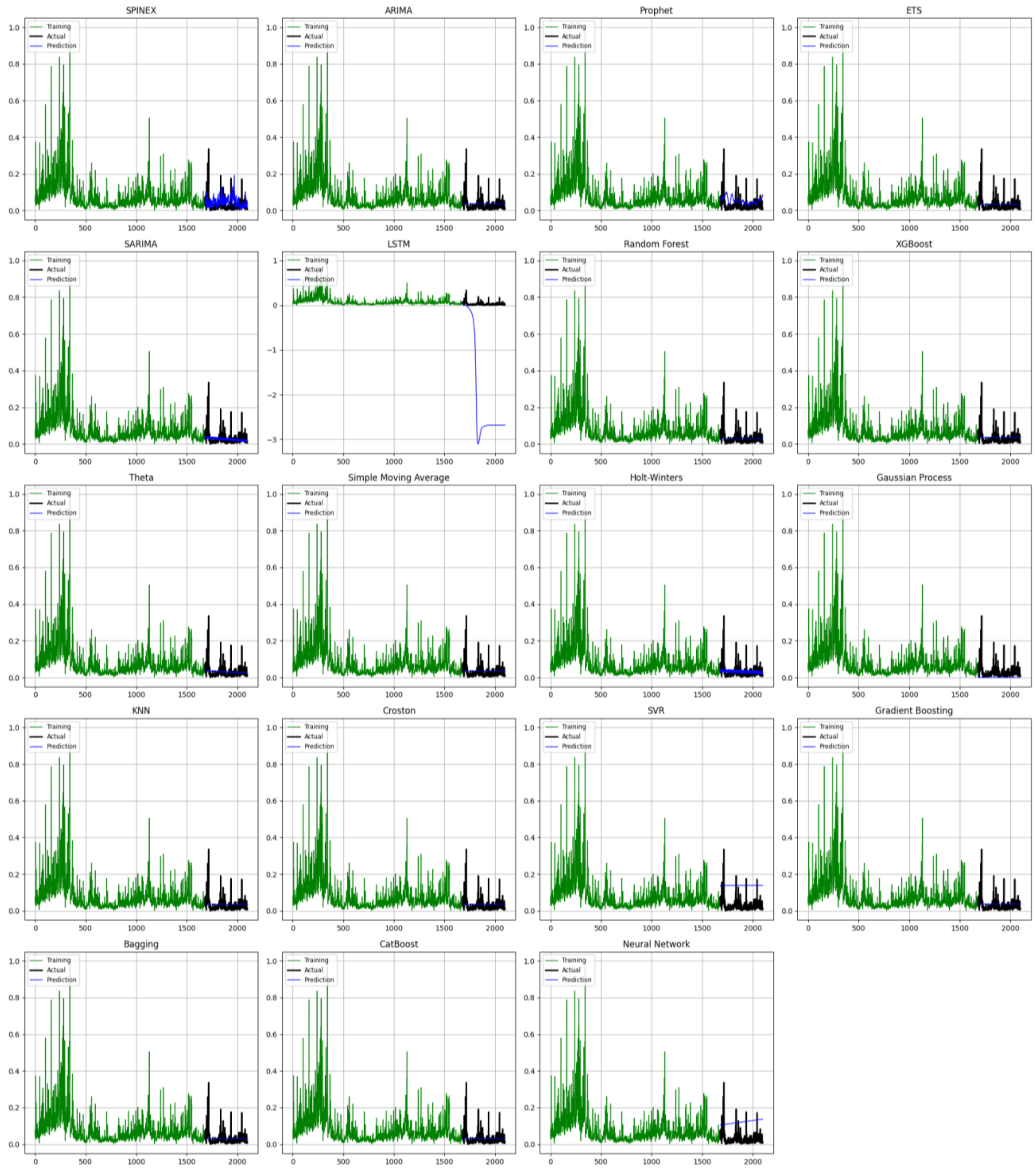

Fig. 6 Visualization of forecasting on Yearly Water Usage dataset (top) and Tata Global Forecasting dataset (bottom)

*4.3 Dataset examination and Pareto efficiency analysis*

Once the above analysis was completed, the same results were examined to identify the most reoccurring complex datasets that received the poorest performance and the most consistently ranked datasets with the best performance across all algorithms. The outcome of this analysis is listed in Table 7. This table shows the top 3 datasets in each category and synthetic and real datasets. Interestingly, all complex datasets are univariates, while those that fall under the consistent datasets have a mixture of both types.

Table 7 Dataset examination

| Type | Complex Dataset | Characteristics | Occurrences | Consistent Dataset | Characteristics | Occurrences |
|---|---|---|---|---|---|---|
| Synthetic data | Dataset 27 | $t_{max}$: 100, $n_{points}$: 5000 | 17 | Dataset 106 | $t_{max}$: 100, $n_{points}$: 50 | 9 |
| | Dataset 96 | $t_{max}$ : 10, $n_{points}$: 5000 | 14 | Dataset 115 | $t_{max}$: 100, $n_{points}$: 50 | 9 |
| | Dataset 99 | $t_{max}$: 100, $n_{points}$: 5000 | 8 | Dataset 91 | $t_{max}$: 1, $n_{points}$: 50 | 9 |
| Real data | Sunspots | 2820/2 | 15 | Yearly Water Usage | 79/2 | 13 |
| | Stock and PM2.5 Prediction | 5650/10 | 14 | Tata Global Forecasting | 2100/8 | 10 |
| | Indian Crime II | 550/2 | 8 | Jaipur | 676/40 | 7 |

A Pareto analysis is performed on synthetic datasets to determine the best-performing time series algorithms based on the selected evaluation metrics (see Table 8). This analysis employs Pareto optimality to identify non-dominated solutions (i.e., those representing optimal trade-offs between different performance metrics: DA, DTW, MASE, and MAD). The concept of Pareto optimality ensures that the final set of recommended algorithms consists of truly superior options, each offering a distinct balance of strengths across various performance criteria. The process starts by normalizing all metrics to a 0-1 scale using min-max normalization. The normalized data is then grouped by algorithm and dataset to calculate mean values for each metric. Each algorithm further aggregates these grouped results to evaluate overall performance across all datasets. Then, a solution is Pareto optimal if no other solution is superior in all metrics simultaneously.

Table 8 Pareto analysis

| Algorithm | Direction Accuracy | DTW | MASE | MAD | Pareto Efficient |
|---|---|---|---|---|---|
| SPINEX | 0.602 | 1.956 | 45.676 | 0.075 | TRUE |
| Holt-Winters | 0.572 | 15.419 | 46.643 | 0.291 | TRUE |
| Theta | 0.550 | 2.451 | 82.481 | 0.100 | TRUE |
| LSTM | 0.502 | 3.756 | 115.265 | 0.090 | TRUE |
| Prophet | 0.546 | 2.058 | 58.910 | 0.095 | FALSE |
| ARIMA | 0.264 | 2.544 | 84.369 | 0.097 | FALSE |
| Croston | 0.000 | 2.602 | 87.962 | 0.101 | FALSE |
| ETS | 0.000 | 2.603 | 87.962 | 0.101 | FALSE |
| Gradient Boosting | 0.000 | 2.689 | 88.144 | 0.101 | FALSE |
| Random Forest | 0.000 | 2.698 | 88.205 | 0.101 | FALSE |
| Bagging | 0.000 | 2.698 | 88.205 | 0.101 | FALSE |
| XGBoost | 0.000 | 2.695 | 88.281 | 0.101 | FALSE |
| Simple Moving Average | 0.000 | 2.669 | 88.733 | 0.101 | FALSE |
| KNN | 0.000 | 2.669 | 88.733 | 0.101 | FALSE |
| CatBoost | 0.000 | 2.666 | 89.402 | 0.101 | FALSE |
| Gaussian Process | 0.255 | 7.354 | 144.260 | 0.097 | FALSE |
| SVR | 0.166 | 4.262 | 157.608 | 0.115 | FALSE |
| Neural Network | 0.518 | 6.952 | 194.656 | 0.110 | FALSE |
| SARIMA | 0.578 | 15.353 | 2625.313 | 0.116 | FALSE |

### *4.4 Complexity analysis*

We evaluate the complexity of the selected algorithms by analyzing the execution time data across across 50, 500, and 5000 samples ($n$). Then, we fit three types of models to this data: polynomial (linear in log-log scale), logarithmic, and exponential. For each algorithm, we compute the regression parameters and the $R^2$ values for these models as a means to measure the goodness of fit and the model with the highest $R^2$ value is considered the best fit, and its corresponding Big O notation is recorded. This analysis reveals that SPINEX demonstrates logarithmic complexity and hence indicates that its execution time scales efficiently with the logarithm of the input size, represented as O(log n), as seen in Fig. 7. It is worth noting that this algorithm was found to have only logarithmic complexity, while others had polynomial or exponential complexity (see Table 9).

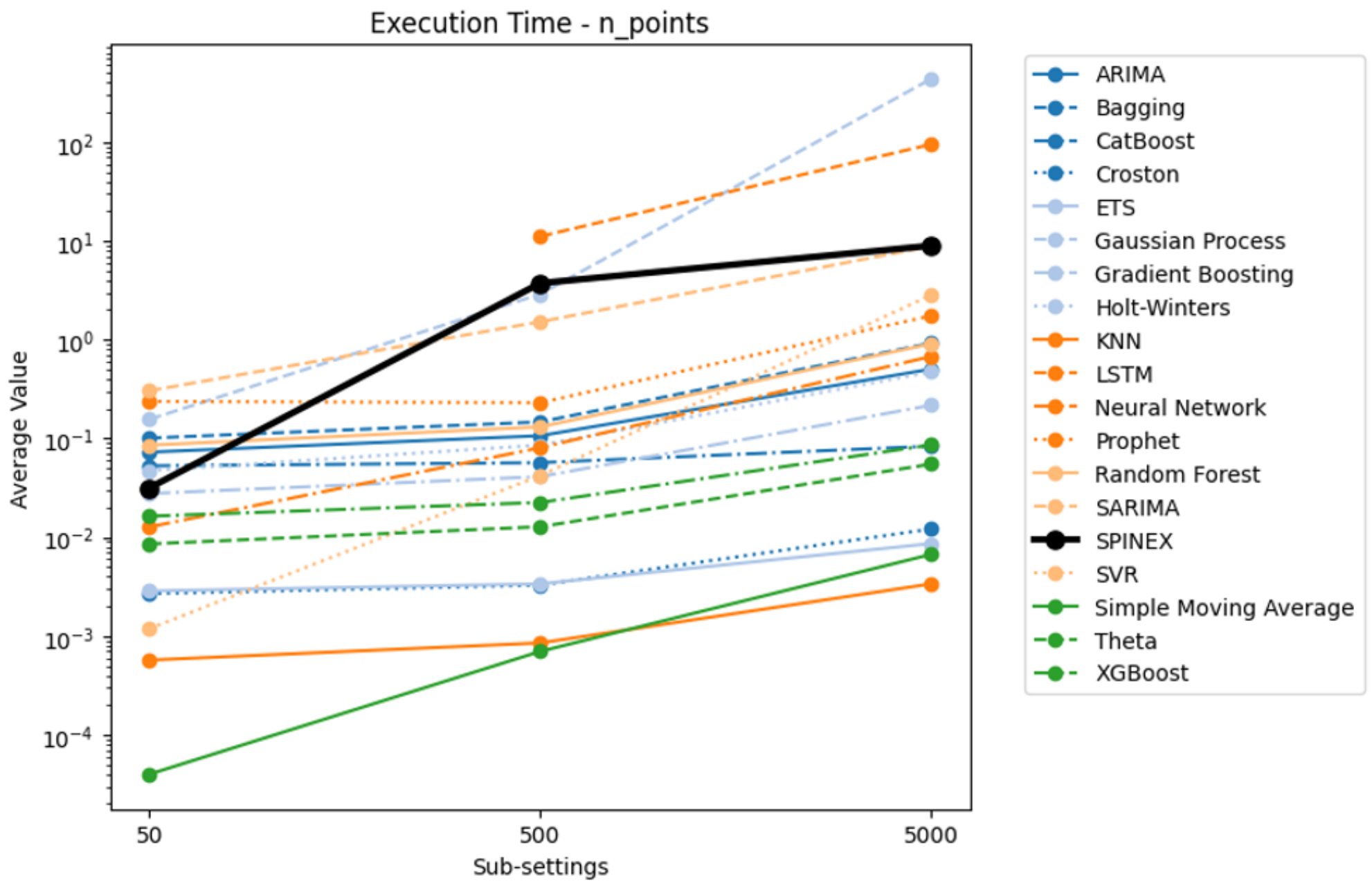

Fig. 7 Outcome of complexity analysis

Table 9 Complexity analysis

| Algorithm | Complexity type | $R^2$ | Big O notation |
|---|---|---|---|
| Prophet | exp | 0.84 | $O(e^n)$ |
| ETS | exp | 0.94 | $O(e^n)$ |
| XGBoost | exp | 0.92 | $O(e^n)$ |
| Theta | exp | 0.85 | $O(e^n)$ |
| Simple Moving Average | exp | 0.98 | $O(e^n)$ |
| Holt-Winters | exp | 0.99 | $O(e^n)$ |
| Croston | exp | 0.89 | $O(e^n)$ |
| Gradient Boosting | exp | 1.00 | $O(e^n)$ |
| SPINEX | log | 0.98 | O(log n) |
| CatBoost | poly | 0.62* | $O(n^{0.12})$ |
| KNN | poly | 0.54* | $O(n^{0.25})$ |
| ARIMA | poly | 0.66* | $O(n^{0.42})$ |
| Bagging | poly | 0.83 | $O(n^{0.46})$ |
| Random Forest | poly | 0.84 | $O(n^{0.49})$ |
| SARIMA | poly | 0.93 | $O(n^{0.71})$ |
| Neural Network | poly | 0.85 | $O(n^{0.86})$ |
| LSTM | poly | 1.00 | $O(n^{0.93})$ |
| SVR | poly | 0.72 | $O(n^{1.36})$ |
| Gaussian Process | poly | 0.97 | $O(n^{1.72})$ |

*note the low value.

*4.5 Explainability analysis*

To showcase the explainability capabilities of SPINEX, two synthetic datasets (No. 1 and No. 11) are provided herein, as taken from two different datasets. Figure 8 shows the predicted segment and three of its neighbors. The same plot also visually represents the neighbors and their overall similarity as compared to the segment at hand. Finally, this plot also shows the individual scores of the similarity measures used to identify them.

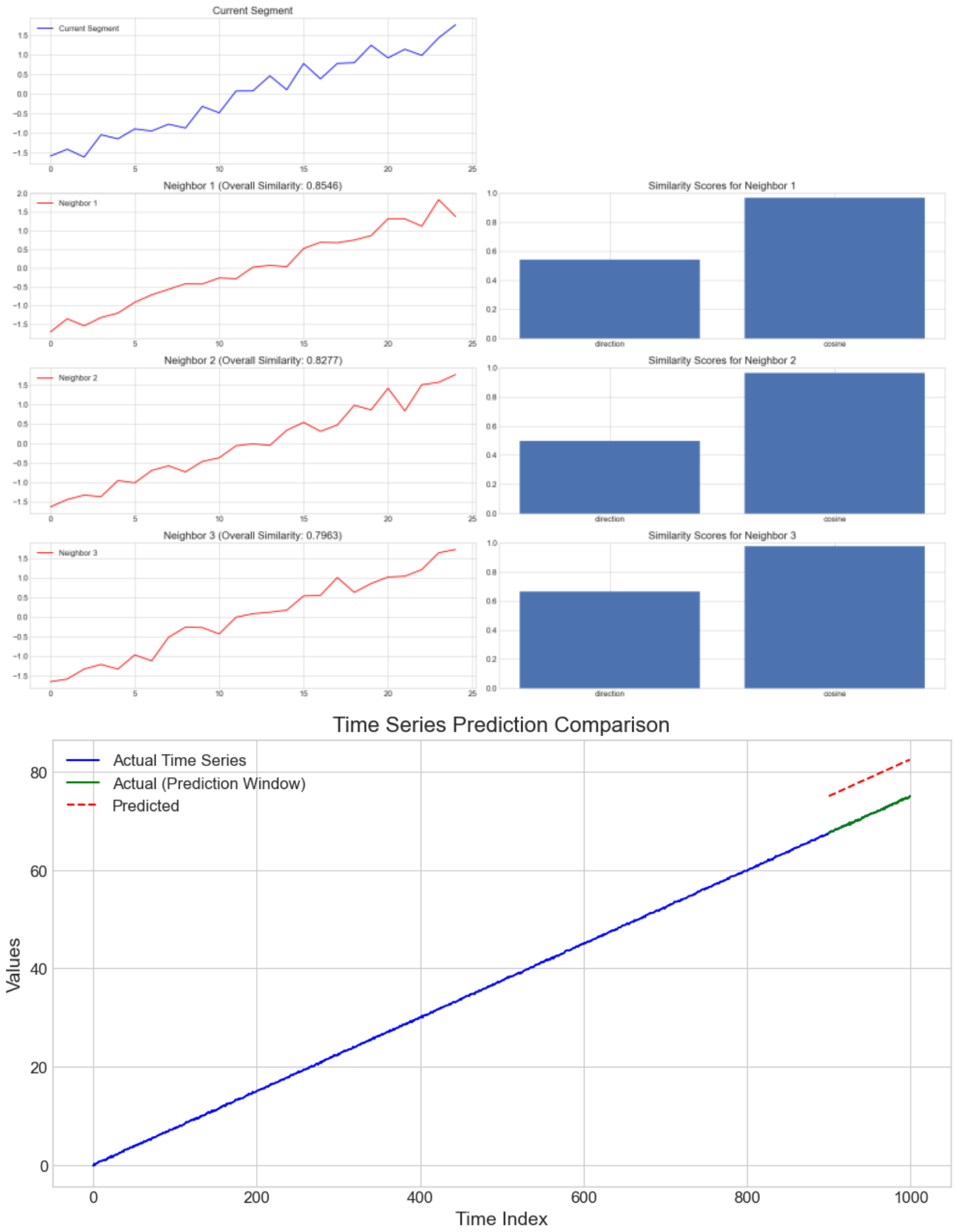

(a) Dataset 1

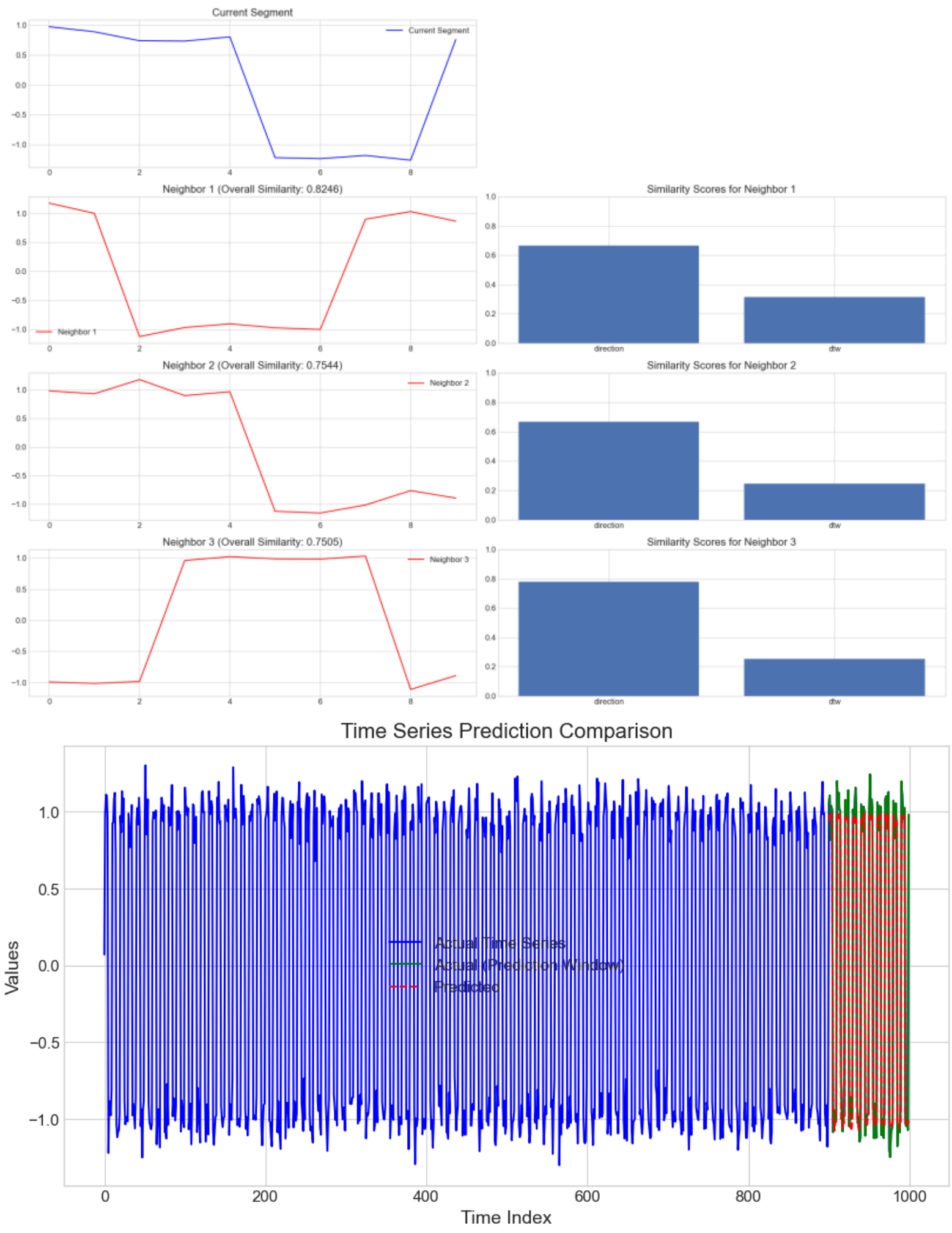

(b) Dataset 11

Fig. 8 Example of explainability capability of SPINEX

## 5.0 Conclusions

This paper introduces a novel member of the SPINEX (Similarity-based Predictions with Explainable Neighbors Exploration) family. This new algorithm enhances time series analysis performance by leveraging the concept of similarity, higher-order temporal interactions across multiple time scales, and explainability. The effectiveness of the proposed SPINEX variant was evaluated through a comprehensive benchmarking study involving 18 time series forecasting algorithms across 49 datasets. Our findings from our experiments indicate that SPINEX consistently ranks within the top-5 best-performing algorithms, showcasing its Pareto efficacy in time series forecasting and pattern recognition while maintaining moderate computational complexity on the order of O(log n). Moreover, the algorithm's explainability features, Pareto efficiency, and medium complexity are demonstrated through detailed visualizations to enhance the prediction and decision-making process.

**Data Availability**

Some or all data, models, or code that support the findings of this study are available from the corresponding author upon reasonable request.

SPINEX can be accessed from [**to be added**].

**Conflict of Interest**

The authors declare no conflict of interest.